\documentclass[prd,reprint,aps,amsmath,amssymb,showpacs,nofootinbib]{revtex4-1}
\pdfoutput=1
\usepackage[utf8x]{inputenc}
\usepackage{graphicx}
\usepackage{enumitem}
\usepackage{slashed}
\usepackage{url}
\usepackage[pdftex]{hyperref}
\usepackage[T1]{fontenc}
\usepackage{bm}
\usepackage{color}
\usepackage{enumerate}
\usepackage{amsmath}
\usepackage{amssymb}
\usepackage{diagbox}
\usepackage{makecell}
\usepackage{subcaption}
\usepackage{mwe}
\usepackage{algorithm}
\usepackage{algpseudocode}

\newcommand{\GeV}{\rm GeV}

\newcommand{\Msun}{{M_\odot}}

\def\apj{ApJ}
\def\apjl{ApJL}
\def\apjs{ApJS}
\def\aj{AJ}
\def\nat{Nature}
\def\mnras{MNRAS}
\def\physrep{physrep}

\def\araa{{Ann.\ Rev.\ Astron.\& Astrophys.\ }}
\def\aap{{A\&A}}
\def\prd{PRD}
\def\jcap{JCAP}

\begin{document}

\title{Exploring the astrophysics of dark atoms}
\author{Akshay Ghalsasi}
\email{ghalsaaa@ucmail.uc.edu}
\affiliation{Department of Physics, University of Cincinnati, Cincinnati, OH 45219, USA}

\author{Matthew McQuinn}
\email{mcquinn@uw.edu}
\affiliation{Department of Astronomy, University of Washington, Seattle, WA 98195, USA}

\date{\today}

\begin{abstract}
A component of the dark matter could consist of two darkly charged particles with a large mass ratio and a massless force carrier. This `atomic' dark sector could behave much like the baryonic sector, cooling and fragmenting down to stellar-mass or smaller scales.  Past studies have shown that cosmic microwave background and large-scale structure constraints rule out $\gtrsim 5\%$ of the dark matter to behave in this manner. However, we show that, even with percent level mass fractions, a dark atomic sector could affect some extragalactic and galactic observables. We track the cooling and merger history of an atomic dark component for much of the interesting parameter space. Unlike the baryons, where stellar feedback (driven by nuclear physics) delays the formation and growth of galaxies, cooling dark atomic gas typically results in disks forming earlier, leaving more time for their destruction via mergers. Rather than disks in Milky Way sized halos, we find the end product is typically spheroidal structures on galactic scales or dark atom fragments distributed on halo scales.  This result contrasts with previous studies, which had assumed that the dark atoms would result in dark disks. Furthermore the dark atoms condense into dense clumps, analogous to how the baryons fragment on solar-mass scales. We estimate the size of these dark clumps, and use these estimates to show that viable atomic dark matter parameter space is ruled out by stellar microlensing, by the half-light radii of ultra-faint dwarf galaxies, and by Milky Way mass-to-light inferences.
\end{abstract}

\pacs{}
\maketitle
\section{Introduction}
Observations of the cosmic microwave background (CMB) and the large-scale distribution of galaxies indicate that only $20\%$ of the matter in the cosmos is baryonic and that the rest is some other substance, termed dark matter \citep[e.g.][]{2016A&A...594A..13P}. The baryonic sector is highly collisional and efficient at radiating energy, resulting in complex dynamics.   Atomic and bremsstrahlung emission allow the baryons to cool and form a disk within Milky Way-sized halos. Further cooling, often by molecules, allows the baryons to lose pressure support and fragment all the way down to stellar masses, with this mass scale determined by how far sound waves are able to travel within a collapse time. Dark matter, on the other hand, is typically imagined to only weakly interact (with itself or the baryonic sector).  As a result, the densest bound structures that dark matter forms, dark matter halos, are a factor of $\sim 100$ more extended than the galaxy disks that reside within them \citep{gunn72, 1996ApJ...462..563N, mo98}. 

This scenario of diffuse, non-interacting dark matter has met with tremendous success at explaining the large-scale structure of the cosmos \citep{2016A&A...594A..13P, 2017MNRAS.470.2617A}, the distribution of matter in galaxy groups and clusters \citep{2017arXiv171003235M}, and limits on the interaction cross section in astrophysical systems \citep{2008ApJ...679.1173R, 2014PhRvD..89b3519D}.  However, studies have claimed that this vanilla dark matter model may not explain certain anomalies on $< 100 ~$kpc scales \citep{1994Natur.370..629M,2011MNRAS.415L..40B, 2011AJ....141..193O}, although see \cite{2014Natur.506..171P, 2016ApJ...827L..23W}.
Motivated by explaining these small-scale anomalies and by the fact that the baryonic sector is so physically rich, a multitude of studies have imagined a more complex dark sector.  Dark matter models with a new dark force that enhances annihilations \citep{2009PhRvD..79a5014A}, large self interactions \citep{2000PhRvL..84.3760S, 2013MNRAS.430...81R, GOLDBERG1986151}, and interactions with baryons \citep{2014PhRvD..89b3519D} have been considered in detail. Several studies have considered the possibility that of `atomic' dark matter, where a component of the dark matter consists of charged particles with an MeV-mass dark `electron'  \cite{2010JCAP...05..021K, 2013PDU.....2..139F, 2014PhRvL.112p1301R, Foot:2014uba}.  This scenario leads to complex cooling physics analogous to the baryonic sector.  \citet{2013PDU.....2..139F} argued that the ``dark atoms'' would cool into a dark disk, leading to a new set of observables.  Such observables include signatures of additional dissipation in observations of galaxy cluster mergers and an unexpected velocity distribution function in direct dark matter detection experiments \citep{2016arXiv161004611A, 2016ApJ...824..116K}.

Various observables have been used to constrain the fraction of the dark matter that could be atomic. Diffuse, darkly charged dark matter is  constrained by self-interaction limits from the Bullet Cluster to be $<30 \%$ of the dark matter \cite{2008ApJ...679.1173R}.  However, if this matter had cooled and fragmented into dense nuggets, the atomic dark matter would behave collisionlessly, avoiding this bound. 
A more robust bound comes from the early Universe.  If the dark atoms were coupled to the thermal bath at any time in the early universe, there would also be a dark CMB. 
 The dark CMB with a temperature today of $T_{d0}$ would drag around the dark atoms, damping the growth of their overdensities and generating acoustic oscillations in their clustering \cite{CyrRacine:2012fz}. CMB and galaxy clustering observations show no evidence for such damping or oscillations. If $100\%$ of the dark matter were atomic, the only viable parameter space has a dark CMB temperature of $\lesssim 0.3~$Kelvin or a dark electron that is more massive than the Standard Model electron \citep{2014PhRvD..89f3517C}.  However, almost all atomic dark matter parameter space is allowed if $\lesssim 5$\% of the dark matter were atomic, as considered here.

This study develops more physical models for the structure formation and astrophysics of dark atoms.  These models show that the assumption of previous studies that dark atoms end up in galaxy-scale disks is rarely justified and, hence, neither are the constraints derived from this assumption.  We do find that such modeling predicts the dark atoms to clump on certain characteristic scales, opening up new avenues for constraining the atomic dark matter parameter space.  We show that observations of dwarf galaxies, galaxy rotation curves, and stellar microlensing may allow percent-level constraints on the fraction of dark matter in some parts of atomic dark matter parameter space.  

Following \citet{2013PDU.....2..139F}, the atomic dark matter models we consider have the following properties
\begin{itemize}
\item The ``dark proton'' is much heavier than the ``dark electron''.   We consider dark proton masses of $1~\GeV < m_X <10~\GeV$, and dark electron masses of $10^{-2}~\GeV < m_c < 10^{-5}~\GeV$.  The subscript $c$ stands for `coolant' as a light dark electron is critical for cooling.   The `dark proton' and `dark electron' only have dark electromagnetic interactions through a massless dark photon $\gamma_{D}$ with the dark fine structure constant $\alpha_{X}$ in the range $10^{-3}-10^{-1}$.  We show that these ranges for $m_c$, $m_X$, and $\alpha_X$ cover much of the interesting parameter space where energy exchange and cooling is possible.\footnote{The upper bound on $\alpha_X$ is chosen to keep dark electromagnetic interactions weak, allowing us to use standard results for atomic processes.}
\item Dark atoms comprise $\epsilon = 5\%$ percent of the dark matter.  We do not expect our qualitative conclusions to change for order of magnitude larger or smaller values.\footnote{The cooling times scale as $\propto {\epsilon}^{-1}$, often quite a bit smaller than the lifetime of the halo. Changing $\epsilon$ would change slightly the regions of our parameter space that can cool efficiently, it would not affect our qualitative conclusions regarding galaxy morphology significantly.} 
\item  The ratio of temperature of dark CMB photons to standard model (SM) CMB photons, $\xi \equiv \frac{T_{d0}}{T_{\gamma 0}} = 0.5$. 
 Current CMB measurements of the effective relativistic degrees of freedom require $\xi \leq 0.5$ \cite{2016A&A...594A..13P}, and the constraints from large-scale structure are somewhat more stringent \cite{2014PhRvD..89f3517C}. Our conclusions are unchanged if instead $\xi \ll 0.5$, except in the relatively small part of parameter space where Compton cooling is the primary coolant.
 \item Finally, we adopt a minimalist model for the dark sector in which there is no feedback on the distribution of dark atoms.  This largely means that there are no dark `supernovae' or something of the like (which would likely be the case if there is no nuclear physics in the dark sector).  We will comment further on this assumption.
\end{itemize}

Note that although we have used dark atoms as an example in this paper, our methodology is applicable to any model where radiative cooling is efficient.

This paper is organized as follows. In Section \ref{sec:cooling}, we will discuss dark atomic cooling processes. We will also recap the cooling processes described in \citet{2013PDU.....2..139F}, and add atomic cooling to the mix. In Section \ref{sec:galformation}, we will discuss hierarchical galaxy formation, and we will describe the rules that determine the halo and galactic-scale distribution of atomic dark matter.  We will then compute the distribution of structure for much of the interesting dark atom parameter space. In Section \ref{sec:darkclumps}, we will discuss the final clump size of the dark atoms, structures analogous to the stellar mass fragments of the baryons called stars.  Finally, Section~\ref{sec:constraints} uses the previous calculations and astrophysical observations to rule out atomic dark matter parameter space. Our calculations assume a $\Lambda$CDM cosmology with $\Omega_{b}\approx 0.04$, $\Omega_{\rm DM}\approx 0.26$, $h=0.71$, $n_s=0.95$ and $\sigma_8=0.8$, consistent with the most recent CMB measurements \citep{2016A&A...594A..13P}.

\section{relevant timescales for the dark atoms}
\label{sec:cooling}
This section calculates the rate at which dark atoms exchange energy, collapse, and can cool.  These rates are crucial for understanding how these atoms condense.  Most of our discussion concentrates on cooling as there are many processes by which dark atoms can cool:  Compton scattering off of the dark CMB, bremsstrahlung, atomic transitions, and even molecular ones.(We consider molecular cooling in a later section.)

The number density of dark protons in a virialized halo at redshift $z$ is given by
\begin{align}
\label{eq:nden}
n_{c}(z) = \frac{ \epsilon \rho_{\rm m,0} \Delta_{\rm halo}(1+z)^{3}}{m_{X}},
\end{align}
where $\epsilon$ is the fraction of dark matter that is in dark atoms, $\rho_{\rm m,0}$ is the present day dark matter density, and $\Delta_{\rm halo}$ is the halo density in units of the critical density (taken here to be the spherical collapse result in Einstein de-Sitter universe, i.e. $\Delta_{\rm halo} = 18\pi^{2}$; \citealt{gunn72}).
An important timescale associated with a virialized halo is it's dynamical time which is the characteristic time needed to fall to the center of a halo (and roughly a tenth the Hubble time). The dynamical time is given by
\begin{align}
\label{eq:tdyn}
t_{\rm dyn} = \left(8\pi G \, \rho_{m} \right)^{-1/2}
\end{align}
where $\rho_{m}$ is the matter density of the system. For a virialized halo at redshift $z$, $\rho_{m} = \rho_{\rm m,0} \Delta_{\rm halo}(1+z)^{3}$. We will need the dynamical time later to compare it to the lifetime of the halo to make sure the halo cools.

Dark atoms that fall into halos likely shock heat to the virial temperature of the halo given by \citep[e.g.][]{BarkanaLoeb2001}
\begin{align}
\label{eq:Tvir}
T_{\rm vir} = \frac{1}{5}\left(\frac{4\pi}{3}\right)^{1/3} GM^{2/3}_{\rm halo}m_{X}^{4/3}{\epsilon^{-1/3}}n_{c}^{1/3},
\end{align}
where $M_{\rm halo}$ is the halo's mass. At these virialization shocks, the dark protons kinetic energy increases first (since they carry most of the momentum).\footnote{A concern is whether this shock heating would happen if the dark matter is less interactive than the baryonic sector. However, even if the Coulomb mean free path is long compared to the system, electromagnetic interactions tend to mediate shocking on considerably smaller scales for the case of standard baryons.  We expect this result will also hold for dark atoms, even for dark parameters that are orders of magnitude away from those for our baryons.} Next, the dark protons, whose entropy is increased by the shock, must exchange energy with the dark electrons. This equilibration happens via particle-particle collisions on a timescale of\footnote{It is possible that collective plasma processes could mediate energy exchange on shorter timescales.}
\begin{align}
t_{\rm equil} &= \frac{m_{X}}{2\sqrt{3\pi}\alpha^{2}_{X}n_{c}} \left(\frac{3T_{\rm vir}}{m_{c}}\right)^{3/2} {\log^{-1}\left(1+\frac{T^{2}_{\rm vir}}{4\alpha^{2}_{X}n^{2/3}_{c}} \right)}.
\end{align}
Since the dark electrons radiate, they must be able to exchange energy with the shock heated protons.  This process is efficient if $t_{\rm equil}$ is smaller than the lifetime of the halo, which we define more specifically later.

Atomic cooling can happen through the processes of Compton cooling, bremsstrahlung and collisional excitation of atomic lines. 
The cooling time -- the time for an average electron to radiate its thermal energy --  for Compton cooling and bremsstrahlung are given by
\begin{align}
\label{eq:bremcomp}
t_{\rm brem}\,&=\,\frac{9}{2^{5}}\left(\frac{3\pi}{2}\right)^{1/2}\frac{m^{3/2}_{c}T^{1/2}}{\alpha^{3}_{X}n(z)x^{2}}; \nonumber\\ 
t_{\rm Comp}\,&=\,\frac{135}{64\pi^{3}x}\frac{m^{3}_{c}}{\alpha^{2}_{X}\left(T_{d0}\left(1+z\right)\right)^{4}},
\end{align} 
where $T_{d0}$ is the temperature of dark CMB today, $x$ is the fraction of unbound dark electrons and we have set the bremsstrahlung Gaunt factor to unity. (For the standard model, the Gaunt factor differs from unity by tens of percent.)   In the next section, we evaluate cooling times at the characteristic densities and temperatures of halos as given by Eqs. \ref{eq:nden} and \ref{eq:Tvir} respectively.

The cooling times are appreciably fast only in the case where $T_{\rm vir}> 0.1 B_X$, where $B_X = \frac{\alpha^{2}_{X}m_{c}}{2}$ is the binding energy of the dark atoms. In this scenario the gas will ionize, whereas in cases where $T_{\rm vir}< 0.1 B_X$, the gas will not ionize and hence not be able to cool.

Another important coolant that has not been included in previous atomic dark matter models is from collisional excitation of atomic lines. The atomic cooling rate of the dark plasma is more complex than bremsstrahlung and Compton cooling and given by Eq.~\ref{gatomic} in Appendix \ref{atomiccool}.

Atomic cooling -- radiation that results from collisions between atoms and electrons -- tends to be important at temperatures within a factor of few from $\sim 0.1 B_{X}$ as at lower temperatures the gas is neutral and at higher temperatures it is highly collisional ionized; only over a relatively narrow range of temperatures is there sufficient density of ions and neutral atoms for this process to be efficient.
At temperatures $T_{\rm vir} \gtrsim B_{X}$, $x(1-x)\approx \frac{2\langle\sigma_{\rm rec} v\rangle}{\langle\sigma_{\rm ion} v\rangle}$. We can use this to estimate the atomic cooling time to be 
\begin{align}
\label{eq:atomic}
t_{\rm atomic}\left(T > B_{X}\right) \simeq \frac{9}{2^{5}}\left(3\pi\right)^{1/2}\frac{T^{2}}{\alpha^{6}_{X}n(z)}.
\end{align}
Comparing this cooling time to that of bremsstrahlung, we find $\frac{t_{\rm atomic}}{t_{\rm brem}} \simeq \left(\frac{T}{B_{X}}\right)^{3/2}$, again noting that $B_X = \alpha_X^2 m_c/2$. Thus, atomic cooling is more efficient for $T_{\rm vir} \sim 0.1 \alpha^{2}_{X}m_{c}$ and less efficient at $T\gtrsim B_{X}$.  

\begin{figure*}
\includegraphics[width=\linewidth]{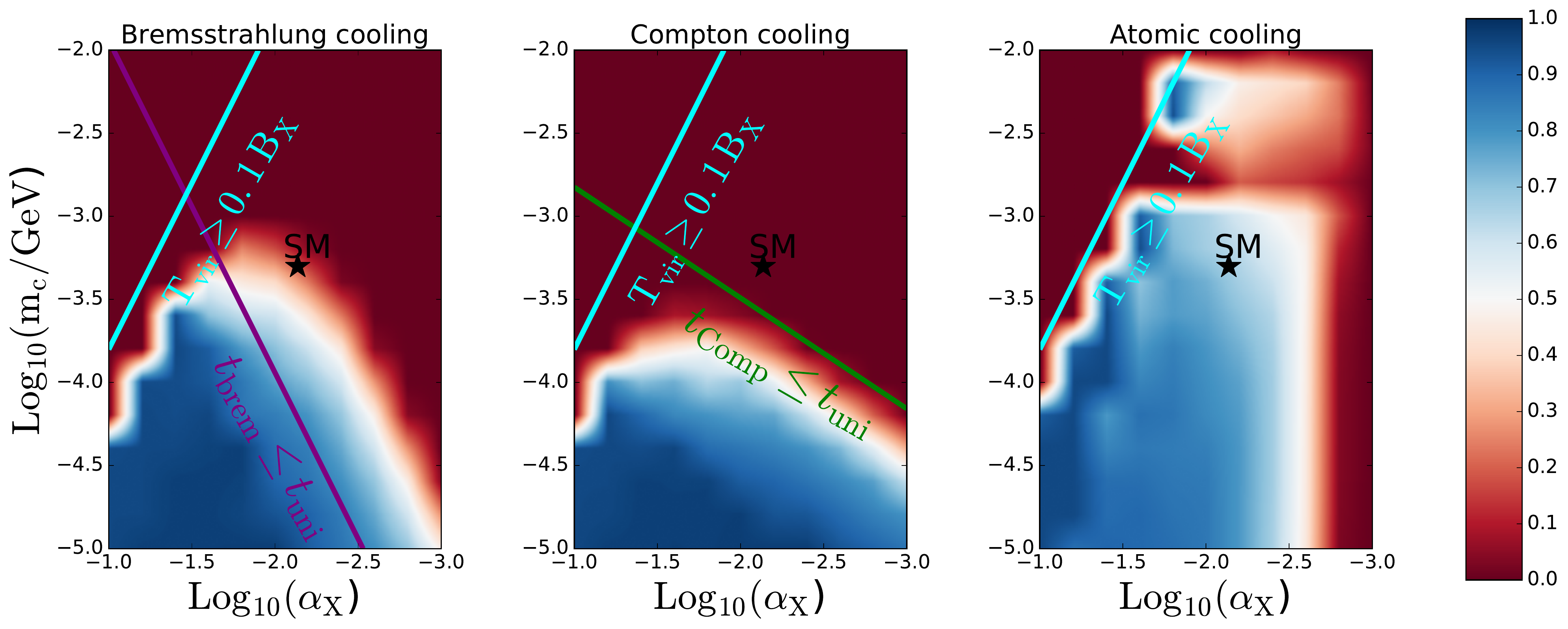}
\caption{Fraction of gas that cools in a Milky Way-like halo ($z=0$ and $10^{12}M_{\odot}$) in calculations with \emph{only} bremsstrahlung cooling, Compton cooling off the dark CMB, or atomic cooling (left, middle, and right panels respectively).  These calculations use the semi-analytic method described in Section~\ref{sec:galformation} that follows the formation of a halo and considers whether relevant timescales are such that the gas can cool.   
Atomic cooling -- which has been omitted in prior studies -- allows more of the dark atom parameter space to cool.  Also shown for comparison are the regions where bremsstrahlung and Compton cooling times are lower that the age of the Universe, $t_{\rm uni}$, calculated assuming a $10^{12}M_{\odot}$ and $z=2$ and where the virial shock likely ionizes the gas ($T_{\rm vir}\geq 0.1 B_{X}$).  These curves were used in \citet{2013PDU.....2..139F} to determine whether the gas cools into a disk.}
\label{fig:diffcooling}
\end{figure*}

Fig.~\ref{fig:diffcooling} shows the fraction of gas that `cooled' -- radiated away its thermal energy -- in a $z=0$ Milky Way-mass halo ($M_{\rm halo} =10^{12}M_\odot$) in a universe with only one cooling mechanism, either bremsstrahlung, Compton, or atomic.   The method used to calculate the fraction of cooled gas is described in the next section, but in short it tracks the formation history of the final halo (such as $T_{\rm vir}$ of its daughter halos) and uses the timescales defined in this section to evaluate whether the gas can cool.   Remember that the interesting parameter space for atomic dark matter models are where cooling occurs.
 Cooling becomes less efficient for lower $\alpha_{X}$ and higher $m_{c}$ for Compton and bremsstrahlung, as expected from Eq.~\ref{eq:bremcomp}. For atomic cooling, the cooling rate from Eq. \ref{eq:atomic} goes as goes as $\alpha^{-6}_{X}$ if $T_{\rm vir} > B_{X}$, which explains the sharp transition into inefficient atomic cooling at low $\alpha_{X}$. Cooling also becomes inefficient when the maximum virial temperature achieved is less than $ 0.1 B_{X}$ i.e. where the dark  atoms never ionize.   Unlike bremsstrahlung and Compton cooling, atomic cooling becomes the dominant coolant as $m_{c}$ increases (see Eqs. \ref{eq:bremcomp},\ref{eq:atomic}). For atomic dark matter with the same masses and fine structure constant as the standard model, it is the most important coolant (with most of the cooling occurring in smaller daughter halos at high redshifts).

Also shown in Fig.~\ref{fig:diffcooling} are the regions where bremsstrahlung and Compton cooling times are lower that the age of the Universe, calculated assuming a $10^{12}M_{\odot}$ at $z=2$, and where the virial shock likely ionizes the gas ($T_{\rm vir}\geq 0.1 B_{X}$).  These curves were used in \citet{2013PDU.....2..139F} to determine whether the gas cools into a disk. Our more detailed calculations show rough consistency with their results, although in the next section we show that most of the cooled gas does not end up in a disk.

\section{Dark Galaxy Formation}
\label{sec:galformation}
Previous work on atomic dark matter has argued that, if this component of the dark matter cools, it naturally forms a galaxy-scale disk \cite{2013PDU.....2..139F}. This conclusion is not obvious. Newly accreted atomic dark matter `gas' tends to shock heat and smoothly populate the entire halo that surrounds a galaxy (which is more than an order of magnitude larger than the size of a galaxy). Cooling then leads to the gas condensing, since it has lost pressure support.  Since free fall to the center of halo takes a somewhat smaller amount of time than collapse, as the gas cools it condenses into the center of the halo, forming a disk whose size is determined largely by angular momentum conservation.\footnote{The time to fall to the center is shorter by a factor of $\sim \epsilon^{1/2}$ than the condensation time.  For this reason, cooling is not in-situ on the scale of a halo, but instead pressure and viscous forces result into the gas condensing into the halo center.  In addition, the large-scale tidal field is responsible for torquing halos and imparting angular momentum.  The amount of angular momentum per unit binding energy only varies by a factor of a few, so all disks and bulges have roughly the same size relative to the halo virial radius \citep{mo98}.}  Finally, the likely end result of cooling is compact gas fragments that behave collisionlessly (as discussed in Section~\ref{sec:darkclumps}), analogous to how baryonic matter fragments into stars.  Over the cosmic history, halos are continually merging and growing.  The central disks inside merging halos often merge themselves, disrupting the disks and making them more spheroidal in shape.  Such disruption is the reason why more massive galaxies than the Milky Way tend to not be disky but instead more spheroidal.  

Astrophysicists have developed fast semi-analytic methods to follow the formation of galactic structures without running expensive numerical simulations \cite{Cole:2000ex, 2015ARA&A..53...51S}. Running such cosmological simulations that cover the large parameter space of atomic dark matter  would be prohibitive, and so these fast methods are crucial for this study.   We describe a basic semi-analytic implementation below, and use it to understand the distribution of atomic dark matter within dark matter halos.
\subsection{Formalism}
Structure formation in our universe has a bottom-up hierarchy.  Smaller dark matter halos form earlier, and these halos go on to merge with others, growing into larger ones.  Thus, the material that constitutes a $z=0$ Milky Way-sized halo, at earlier times resided in smaller halos. The extended Press-Schechter formalism provides an analytic method to follow a halo's merger history \cite{press, bond91, lacey}.  This formalism reproduces the statistics of halo merger histories seen in fully cosmological simulations \cite{2008MNRAS.386..577F}.

In the extended Press-Schechter formalism, a halo with mass $M_{\rm halo} = M$ is considered to have formed at a redshift of $z$ if the $z=0$ linear overdensity averaged over a spherical region of mass scale $M$,  $\delta_M$, exceeds
\begin{align}	
\omega(z) = 1.69/D(z).
\label{eqn:deltac}
\end{align}	
where $D(z)$ is the linear growth factor for the matter overdensity and $\omega(z)$ is the critical linear theory overdensity \citep{gunn72}. \footnote{Eq.~\ref{eqn:deltac} is exact for a matter-dominated Universe where $D(z) = a$, where $a$ is the scale factor.  This equation is also a good approximation for the $\Lambda$CDM cosmology.  For this study we use $D(z)$ for matter-dominated universe, a simplification that does not affect our conclusions appreciably.}  (Note that this model is defined in Lagrangian space, where there is a one-to-one relation between $M$ and the region's Lagrangian radius.)   The largest mass scale for which $\delta_M >\omega(z)$ sets the halo's mass in this model.  At higher redshifts, $D(z)$ is smaller and, hence, $\delta_M >\omega(z)$ for the same region is likely to be satisfied at smaller $M$ (as the RMS of $\delta_M$ decreases monotonically with $M$), reproducing the bottom-up hierarchy of structure formation.  This algorithm allows one to associate a halo of mass $M$ with the halos that constituted it at an earlier time.  Note that the statistics of $\delta_M$ are Gaussian, which allows for straightforward analytic formula in terms of the RMS of $\delta_M$.   See Appendix \ref{mergeralgo} for details.

While the merger tree provides the history of the collisionless component of the dark matter, to understand the condensation of the atomic dark matter, the rules governing this component are followed on top of the merger tree.  In astrophysics, similar calculations are done to follow galaxy formation, and they are called `semi-analytic galaxy formation models'.  The exact algorithm used for this computation is described in detail in Appendix \ref{mergeralgo}.  Here we briefly sketch the approach.  To add galaxy formation to this tree, we start at the end branches and descend towards the trunk.\footnote{After the initial time ($z=10$ in our calculations), the tips of the merger tree branches are set by the mass resolution of the merger tree.}  Every time a merger occurs, we determine whether the atomic dark matter in the merged halo can cool by comparing its cooling time, dynamical time, and equilibrium time ($t_{\rm cool}$, $t_{\rm dyn}$ and $t_{\rm equil}$) to the halo lifetime ($t_{\rm halo}$)\footnote{Mass of halos in which dark atoms are able to cool was also explored in \cite{Buckley:2017ttd}.}, defined as the time for the current halo mass to double through subsequent mergers and accretion.  If $t_{\rm cool}, ~t_{\rm dyn},~ t_{\rm equil}$  < $t_{\rm halo}$, we consider the entire reservoir of dark baryons to have cooled.  When the gas within a halo cools, it forms a central disk (because of angular momentum conservation). We further assume that this disk has fragmented into small clumps, which will happen unless there is a strong source of feedback.  (More on feedback later as well as justification for our assumption that it is likely unimportant for atomic dark matter.) At the next time step, a halo may merge with other halos.  The central disks may be massive enough to spiral together by dynamical friction (see Eq. \ref{eq:tmerge} and preceding discussion of dynamical friction). If these central disks are comparably massive, the gravitational torques during their collision will destroy the disks, making the mass distribution spheroidal (`a bulge') with an extent comparable to the extent of the disk. (The clumps of dark atoms behave collisionlessly and so there is no way to reform a disk after a significant gravitational interaction.)\footnote{In our algorithm, the disk is destroyed if ${M_{2}}/{M_{1}} > f_{\rm disr}$, where $M_{1}$ and $M_{2}$ are the masses of the merging galaxies  and $M_{1} > M_{2}$. We choose $f_{\rm disr} = 0.5$, but values of $0.3\leq f_{\rm disr} \leq 0.5$ have been used in literature. The results do not change appreciably for values of $f_{\rm disr}$ of 0.3 and 0.5.}  If there is a significant mass ratio between the merging disks, the larger disk will survive the merger and the smaller disk will end up contributing to the bulge component. If some gas is uncooled in the halos prior to merging, the new conditions of the merged system (e.g. $T_{\rm vir}$ and $t_{\rm halo}$) may allow cooling in the merged system.\footnote{We ignore the torques between the baryonic disk and the dark atom disk, which would require detailed modeling of the baryons and their feedback.  Since atomic dark disks form relatively late, as discussed later, often the dark atom disks are more massive than their baryonic counterpart for our models with $\epsilon =0.05$, which might justify ignoring such torques.  We note that including torques from the baryons will only further act to reduce the (small) dark disk fractions in our model.}

We track for each halo the fraction of gas that is uncooled, that is cooled, that is cooled in disks, or that is cooled into bulges.  After each time step, these quantities are combined and evolved with specific, physically motivated rules. For the full algorithm, see Appendix~\ref{mergeralgo}.

To give some intuition into whether the dark atoms form a disk or more spheroidal structures, consider the following scenarios:
\begin{description}
\item [Scenario 1 (No cooling)]  In the simplest scenario, the dark atoms never cool.  The gas stays in a virialized state, filling the entire DM halo.  There is no bulge or disk.
\item[Scenario 2 (Dark Bulge)]  
This scenario occurs when cooling is efficient.  At some high redshift, the dark atoms cool in small halos.  Since cooling is a runaway process, accelerating as gas gets denser, the likely end product of cooling is a disk of gravitationally bound objects (`dark clumps').\footnote{These clumps are the dark analogue to stars.  In the baryonic case, stellar feedback processes, such as supernovae, suppress much of the gas from becoming stars.  Minimal atomic dark matter scenarios likely do not have feedback processes.}  This disk and its more diffuse dark matter halo proceed to merge with other halos.  Upon merger with another halo, the less massive of the daughter halos spirals to the center of the more massive halo by dynamical friction (see Eq. \ref{eq:tmerge}). If galaxies of comparable mass in those halos merge (a `major merger'), the dark disks are disrupted and the cooled baryons form a spheroidal bulge. Multiple minor mergers can also give an appreciable bulge fraction.
\item [Scenario 3 (Disk formation)] A disk results only under special circumstances where the gas is able to cool but where cooling is relatively inefficient so that it occurs late, after the most massive progenitor halo-- the `main progenitor,' abbreviated MP --  has had its last major merger. We find that only a small region of the parameter space forms a significant disk.
\item [Scenario 4 (Dark halo)] Finally, if cooling occurs in small halos and these objects merge with much larger halos so that the timescale for dynamical friction is long, then the dark clumps will stay in the outskirts of the halo.  This scenario is analogous to the galaxies that orbit the potential of galaxy clusters.
\end{description}
In reality, all three components (halo, bulge and disk) can coexist and our calculations allow for this possibility.

\subsection{Merger tree evaluation for SM parameters}
\begin{figure*}
\includegraphics[width=\textwidth]{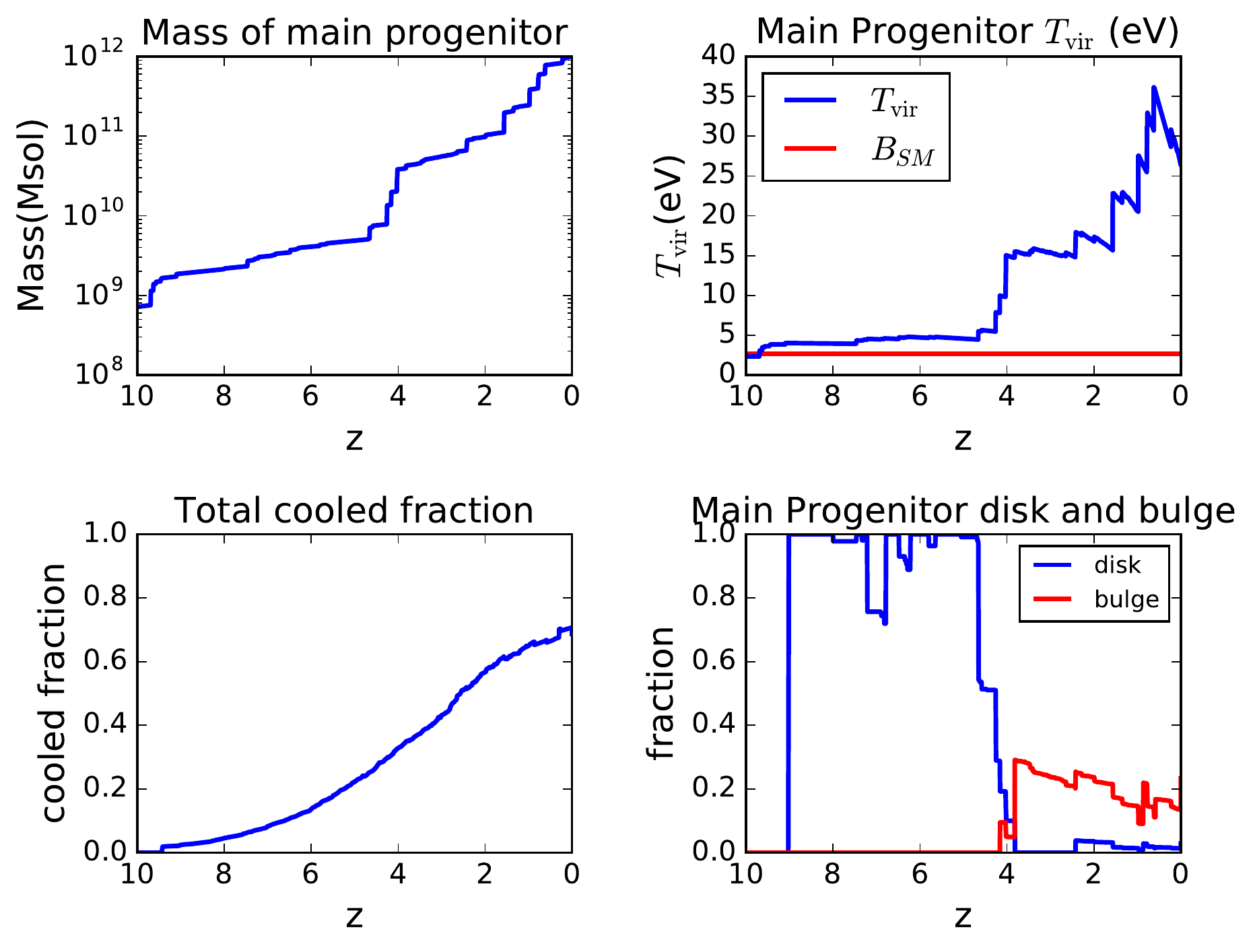}
\caption{Properties of the atomic dark matter in the most massive progenitor halo as a function of redshift for a $z=0$ halo with a Milky Way-like mass of $M=10^{12}\Msun$, for the case where the atomic dark matter has the same masses and fine structure constant as the Standard Model baryons. The top left panel shows the mass of the main progenitor as a function of redshift.   The top right panel shows the virial temperature of the main progenitor and the red line shows $B_{SM} = \alpha^2 m_e/2$, somewhat above the temperature at which the gas becomes ionized and hence can cool. Bottom left panel shows the dark atom mass fraction that has cooled as a function of redshift, and the bottom right panel shows the mass fraction in the main progenitor that resides in a bulge or the disk; most of the cooled atomic dark matter was in disks at high redshifts but now resides in the halo.}
\label{fig:mpsm}
\end{figure*}

To develop some intuition, first consider the case where the dark atoms have Standard Model (SM) parameters: $\alpha_{X} = 1/137,~m_{c} = 5\times 10^{-4}~\GeV,~ m_{X}=1~\GeV$.  The halo mass resolution we adopt in our merger tree for these and subsequent calculations is $M_{\rm res} = 3\times 10^{7}M_{\odot}$, well below the $10^{12}M_\odot$ mass of the Milky Way.  Our results do not change appreciably if we take $M_{\rm res} = 10^{8}M_{\odot}$. The top panels in Fig.~\ref{fig:mpsm} show the mass and the virial temperature of the main progenitor (MP) as a function of redshift. The bottom-left panel shows the mass fraction of gas that has cooled as a function of redshift, showing that $\sim 70 \%$ of the gas has cooled by the present. 

The bottom-right panel in Fig. \ref{fig:mpsm} also shows the fraction of the total mass in the disk and bulge components of the MP, again for the case of SM parameters.  At $z = 0$, most of the baryons form a bulge component and there is a negligible fraction in the disk.  Following the history of the MP, its cools to form a disk at $z \approx 10$, but then undergoes a major merger at $z\approx 4$ destroying its disk component. Subsequent mergers then contribute mostly to the bulge (except for gas that cools which contributes to the disk). As smaller systems merge with the MP, most of their atomic dark matter ends up in the virialized halo because the dynamical friction time is not short enough to reach the center of the MP. 

 It may seem contradictory that the dark baryons with SM parameters do not form a disk since SM baryons do end up forming a disk for Milky Way-like halos. However, we note that the SM case for dark atoms should not reproduce the properties of observed galaxies because our dark atoms do not have nuclear physics that feedback in the form of additional radiation and supernovae.  Such feedback plays a significant role in recycling baryonic gas that make it into a galactic disk back into the diffuse dark matter halo.  For example, even though the cooling times are less than the age of the Universe, only 20$\%$ of SM baryons associated with the Milky Way halo condense into the Galaxy, and this fraction is even less in both smaller and larger halos than the Milky Way's halo \citep{2010ApJ...717..379B}.

\subsection{Scanning over the parameter space}
Now that we have considered the specific case of SM parameters, let us consider a broader range of $m_{c},\, \alpha_{X}, \, m_{X}$, focusing approximately on the interesting range where cooling can happen. For the proton masses we consider, $m_{X} = 1~\GeV$ and $10~\GeV$. For dark electron masses we consider $m_{c}\subset \left[10^{-2}~\GeV, 10^{-5}~\GeV\right]$ and fine structure constants of $\alpha_{X}\subset \left[10^{-1},10^{-3}\right]$. The upper bound on $\alpha_{X}$ has been selected for the electron to remain non-relativistic in dark atoms, while the upper bound on $m_{c}$ is chosen because above this value the gas never becomes ionized for the virial temperatures associated with our choices of $m_{X}$ and, hence, likely cannot cool.  Values of $m_{c}$ lower  than $10^{-5}~\GeV$ result in the gas being photoionized by the dark CMB radiation today, which complicates our calculations (as it eliminates atomic cooling, makes Compton processes heat the gas for $T< T_{d0}$, and adds additional photoheating).

\begin{figure*}
\includegraphics[width=\linewidth]{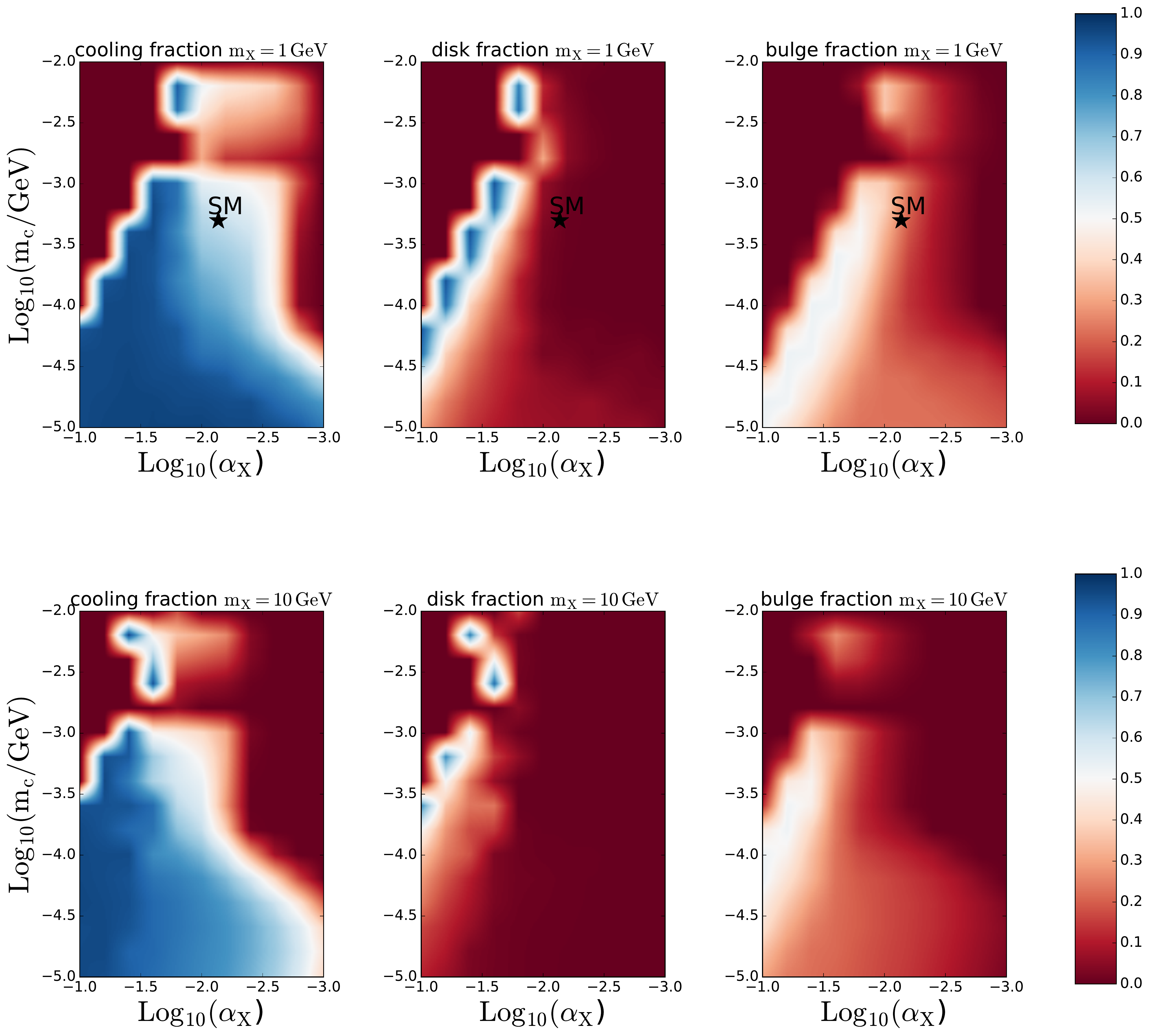}
\caption{Fraction of atomic dark matter that has cooled (left panels), that has formed a disk (middle panels), and that has formed a bulge (right panels) as a function of dark fine structure constant, $\alpha_{X}$, and dark electron mass, $m_{c}$.  We have fixed the dark proton mass at $m_{X}= 1~\GeV$ (top panels) and $m_{X}=10~\GeV$ (bottom panels) and all calculations assume that $5\%$ of the dark matter is atomic.  Dark disks only forms in a small region of parameter space.  Note that the ridges are an artifact of the grid spacing used for these calculations.}
\label{fig:pscan}
\end{figure*}

Figure \ref{fig:pscan} plots the cooled fraction (left panels), the disk fraction (middle panels), and the bulge fraction (right panels) of a $z=0$ Milky Way-like halo, varying $\alpha_{X}$ and $m_{c}$ over the specified ranges and all calculations use our fiducial $\epsilon=0.05$. Note that the `graininess' of the plot is due to linear interpolation between values  calculated for a grid of points on a $\rm 11 \times 16$ grid, corresponding to a step size of 0.2 in $\rm log_{10} m_{c}$ and $\rm log_{10}\alpha_{X}$. These calculations average the results of five different merger trees.

The top panels in Fig.~\ref{fig:pscan} show models with a dark proton mass of $m_{X}=1~\GeV$. For the largest $\alpha_{X}$ and $m_{c}$ shown in these panels, the temperature never goes above $0.1 \, B_{X}$ and, hence, the dark gas is unable to cool.  An appreciable fraction of dark atoms end up in a disk around the line given by $T = B_X$ for the final halo (such that cooling occurs late). As can be seen the total cooled fraction decreases with decreasing $\alpha_{X}$ or increasing $m_{c}$ as expected from the cooling rates described in Section \ref{sec:cooling}. Only $\mathcal{O}(10\%)$ parameter space forms an appreciable disk. However a large part of the parameter space forms an appreciable bulge fraction.

The bottom panels in Fig.~\ref{fig:pscan} show that many of the same trends hold for $m_{X} = 10~\GeV$ case.  One exception is that cooling is less efficient owing to the larger virial temperatures, longer $t_{\rm equil}$, and lower density of coolants for fixed $\epsilon$. On the other hand higher virial temperature allow ionization and hence cooling for higher values of $m_{c}$ compared to $m_{X} = 1~\GeV$ case.   We note that, for the considered range of $\alpha_X$ and $m_c$, a dark proton with $m_{X} = 0.1~\GeV$ would result in smaller virial temperatures for Milky Way-like halos, such that cooling would occur in only a small fraction of the parameter space.

\section{Mass of Dark Stars}
\label{sec:darkclumps}

This section estimates the characteristic mass of dark atom fragments, which we use in the next section to constrain dark atom models.  These fragments are the end product of cooling, analogous to stars in the baryonic sector.   Assuming the dark proton and electron are fermions, these fragments are likely to be the dark analog of white dwarfs (although the timescale to radiate energy and reach this limit could be long), or, if they are above the dark atom's Chandrasekhar mass of $\approx 1.4$~(GeV/$m_X$)$^{1/2}M_\odot$, they are are likely to be black holes.

 In the baryonic sector, observations show that stars show a characteristic mass of a couple tenths that of the Sun, with most of the total mass in stars within a factor of few of this scale, and with a power-law tail in number density to higher masses with index $-2.3$ \citep{2010ARA&A..48..339B}. This mass distribution holds over environments enriched with a broad range of metallicities, although it is believed that stars that form in unenriched regions were more massive.
Explaining the characteristic mass of stars is a topic of ongoing theoretical and numerical work, but there is a basic picture for how the characteristic mass of stars comes about \citep{1976MNRAS.176..483R, 1977ApJ...211..638S, 2005MNRAS.359..211L, krumholz11, 2016MNRAS.458..673G}.  The picture is of gas fragmenting on smaller and smaller scales as it cools, loses pressure support, and condenses. Namely, diffuse cooling gas tends to stay isothermal, and as isothermal gas gets denser, sound waves can communicate over increasingly less massive regions in a gravitational collapse time ($t_{\rm dyn}$).  The length and mass scale over which sound waves can travel in a dynamical time is called the Jeans radius and Jeans mass and given respectively by
\begin{align}
\label{eq:Jeansmass}
R_{J} = \left(\frac{15 T}{4\pi G m^{2}_{X} n_{X}}\right)^{1/2} {\rm ~~~ and~~~} M_J = \frac{4 \pi}{3} m_{X} n_{X} R_J^3,
\end{align}
where $n_X$ is the dark proton density.  Eventually, the gas becomes dense enough that it is no longer able to radiate its energy sufficiently and heats up, halting fragmentation.  The stellar mass scale arises from evaluating Eq.~\ref{eq:Jeansmass} at the applicable temperatures and densities.\footnote{For the baryonic gas in our universe, deuterium burning and dust formation can complicate this simple picture, wrinkles that do not apply to our atomic dark matter fragments.}

Theoretical calculations for the size of these cores in metal enriched environments tend to be $M_\odot \sim 10^{-2} -10^{-1}M_\odot$  \cite{1976MNRAS.176..483R, 1977ApJ...211..638S, 2005MNRAS.359..211L}, somewhat smaller than the characteristic mass of stars.\footnote{Small protostars may continue to accrete an ${\cal O}(10)$ factor in gas because the acoustic contact at an earlier time was over a larger region such that the gas was able to self organize over larger masses than the minimum fragmentation mass \cite{1977ApJ...214..488S}.} Calculations show that these cores are larger with $M_J\sim 500~M_\odot$ \citep{2002ApJ...564...23B} in environments unenriched by stellar nucleosynthesis because the gas radiates less efficiently and, hence, does not reach as low temperatures.  For more massive protostars, stellar radiation likely halts accretion:  For unenriched environments, calculations predict that this occurs at somewhat smaller masses than the Jeans mass, $M\sim 100~M_\odot$ \citep{2002ApJ...564...23B, 2004ARA&A..42...79B}.  However, without fusion power -- the limit that we assume applies for atomic dark matter -- the characteristic mass of `dark clumps' would likely be closer to $M_{J}$.  (Since our atomic matter does not have nuclear physics, we do not call the clumps `stars'.)  

Our approach is similar to that developed for baryons.  We will evaluate the Jeans Mass, Eq.~\ref{eq:Jeansmass}, at the temperature and density where gas cannot cool efficiently.  
For these calculations, we assume that the cooling radiation of dark fragments is insufficient to affect surrounding gas and its fragmentation; such feedback does happen for standard baryons.  However, because of nuclear fusion, radiation is much more important for baryons than if they were powered solely by gravitational energy:   For a solar mass star, there is more than a hundred times more energy to fuse the hydrogen into helium than the gravitational energy to collapse to a degeneracy pressure supported white dwarf.
Thus, it is likely that radiative feedback can be neglected, particularly in the parameter space where the dark clumps are less massive than stars.  In addition, whether dark fragments even radiate ionizing and dissociating photons depends on the temperature of their atmospheres and how it compares to the binding energy of dark atoms, $\alpha_{X}^2 m_c$.  With no internal energy generation, it is likely that a dark fragment's surface has a low temperature as it collapses.\footnote{Even if there were some internal energy generation, such as from dark matter annihilations, the unimportance of radiation backgrounds likely still holds for much of our parameter space. For baryonic matter, \citet{carter74} notes that it is even a conspiracy that stars like the Sun exist and slightly smaller $\alpha$ would result in all stars being dominated by convective energy transport and hence red.  In addition, even if the dark clumps radiate ionizing photons, photoionization tends to be a relatively weak form of feedback that has trouble halting ionized gas from cooling.} 
\subsection{dark atomic cooling}
Let us begin by assuming that the gas can cool only via atomic cooling and that there are no molecules present.  We discuss the case with molecules in the next section.  In the case of atomic cooling, we expect regions where gas can cool efficiently at halo densities will allow the gas to cool to $T \sim 0.1 B_{X}$, reminding you that $B_{X}$ is the binding energy of dark hydrogen.  Both bremsstrahlung and atomic cooling become increasingly efficient with decreasing temperature, driving the gas to lower and lower temperatures.  However, below $T \sim 0.1 B_{X}$, their are insufficient dark electrons in the Maxwellian tail to excite line cooling and also the gas starts to become neutral, which further shuts off all cooling processes.  The exact temperature the gas cools to depends on $n$, $m_c$, and $\alpha_{X}$, but the prefactor of $0.1$ is rather generic owing the strong exponential dependence both of the ionized fraction and collisional cooling on temperature.

One of the ways cooling can become inefficient is if the gas reaches thermal equilibrium. However, it is difficult for the lowest allowed atomic transitions in the gas to reach thermal equilibrium because the transition times are short, with atomic hydrogen's spontaneous transition rate from the $2p$ to $1s$ being roughly $A_{21} \sim 10^{9}$~s$^{-1}$, scaling as $\alpha_{X}^5 m_c$.  The number density at which collisional excitations become equal to spontaneous decays so that dark atomic transitions go into thermal equilibrium is
\begin{equation}
\label{eq:thermalequilbrium}
n_{\rm TE} = 3\times 10^{19} {\rm cm}^{-3} \times \left(\frac{\alpha_{X}}{\alpha} \right)^{6} \left(\frac{m_c}{m_e} \right)^{3},
\end{equation}
where we have assumed a collision rate of $\Gamma_{\rm coll} =  n_X \pi \alpha_{X} a_0^2 f$  at a temperature of $T = \frac{B_{X}}{10}$, where $a_0$ is the `Bohr radius'. Here $f \sim 10^{-3}$ is the exponentially small fraction of electrons in the Maxwellian tail that can collisionally excite transitions. We find that $n_{\rm TE} $ is not achieved in any of our parameter space.
When transitions are not in thermal equilibrium, the gas continues to cool, condense and fragment, unless the gas becomes optically thick to a process that destroys cooling lines. The absorption process that is most likely to set the opacity is free-free absorption (the inverse process of bremsstrahlung). For the free free optical depth to be unity for a cloud of size the Jeans radius (Eq. \ref{eq:Jeansmass}) at the Ly$\alpha$ line for hydrogen -- the major atomic cooling line -- we find

\begin{align}
\label{eq:ff}
n_{\rm ff} = 7\times 10^{9} {\rm cm}^{-3} \left(\frac{\alpha_{X}}{\alpha} \right)^{2}  \left(\frac{m_c}{m_e} \right)^{3}\left(\frac{m_X}{m_p} \right)^{2/3}
\end{align}
where we have assumed a temperature of $T = 0.1 B_{X}$.\footnote{If we evaluated at $\nu \sim T$ characteristic of free-free emission, we would find two orders of magnitude lower densities (and one order of magnitude lower dark clump masses).  However, free free emission is much more likely to be in thermal equilibrium and, hence, surface emission because equilibrium is established by the sea of Coulomb interactions.}
This gives us the scaling of the Jeans mass due to final density set by free-free opacity as
\begin{align}
\label{eq:mjff}
M^{\rm ff}_{J} \simeq 1800 M_{\odot} \times \left(\frac{\alpha_{X}}{\alpha} \right)^{2}\left(\frac{m_X}{m_p} \right)^{-7/3}
\end{align}

For values of $\alpha_{X}$ larger than the fine structure constant, it is possible that Compton and double Compton scatterings destroy the Ly$\alpha$ line more efficiently (for smaller densities) than free free absorption. However it is not obvious whether thermalization proceeds efficiently at these densities. We ignore this possibility here.  

Finally, we ignore two other thermalization processes:  bound-free absorption and H$^-$ absorption.  Both of these processes are extremely temperature dependent and so their effects on opacity are difficult to calculate.  Bound-free needs to be from electrons in the $n=2$ state, which are suppressed by density (with fraction equaling $n/n_{\rm crit} \exp[-\Delta E/T]$) and by ionization.  Furthermore, unlike free-free, both processes tends to be narrow band owing to the photoionization cross sections generically scaling as $\nu^{-3}$; this allows the gas to cool at other wavelengths or, if such cooling is sufficiently blocked, to heat up and collisionally ionize these absorbers.  Furthermore, these absorbers are more sensitive to dissociating ionizing backgrounds.

\begin{figure*}
\includegraphics[width=\linewidth]{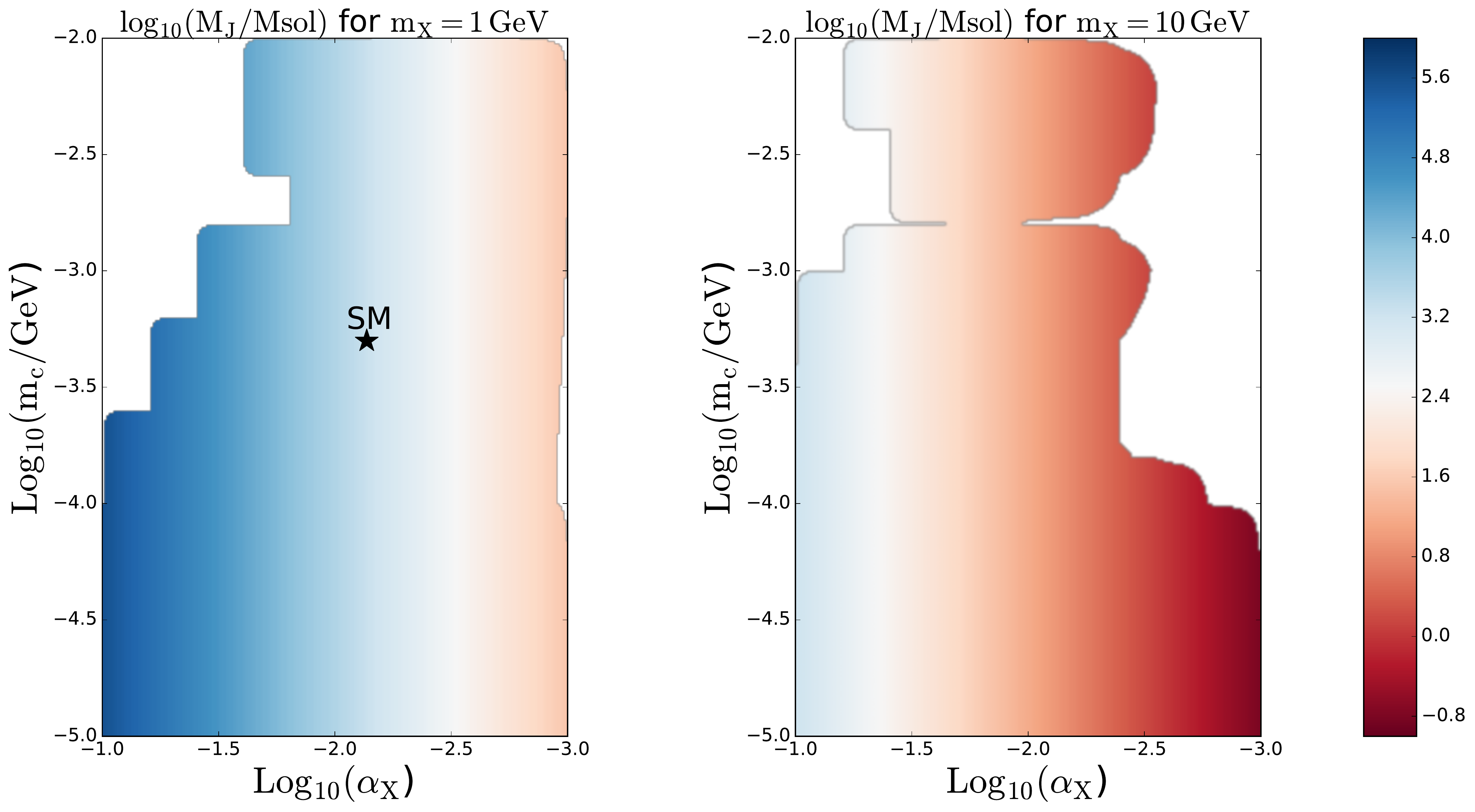}
\caption{Estimated mass of dark clumps for atomic cooling case with $\epsilon=0.05$, $m_{X}=1~\GeV$ or $m_{X}=10~\GeV$. These estimates assume that the mass of clumps is determined by the Jeans mass at a density set by when free-free absorption prohibits atomic line cooling and a temperature characteristic of ionized gas of $0.1 \alpha_{X}^2 m_c$. We only consider parameter space where the gas has cooled in a $10^{12}\Msun$ halo.}
\label{fig:jeans}
\end{figure*}

Figure~\ref{fig:jeans} shows the estimated mass of stars, estimated using $M_{J}$ for $T = \alpha^{2}_{X}m_{c}/10$ and evaluated for when cooling becomes inefficient owing to free-free absorption, $n_{\rm ff}$. With the $m_{c}$ dependence of $M^{ff}_{J}$ in Eq. \ref{eq:mjff}, Jeans mass only depends on $\alpha^{2}_{X}$ and the higher the $\alpha_{X}$ the higher the mass. The Jeans mass also decreases with increasing $m_{X}$.

\subsection{dark molecular cooling}
Dark molecules, a bound state between two dark hydrogen atoms, could potentially form and enable to gas to cool to much lower temperatures than atomic cooling.  Furthermore, molecules reach thermal equilibrium at much lower densities owing to their smaller spontaneous emission coefficient relative to atoms.  These differences result in a gas of dark molecules fragmenting on different scales than one of atoms.  

However, it is very difficult to predict whether dark molecules will form.  In our Universe, molecular hydrogen formation is typically highly out of thermal equilibrium. (At low redshifts, most forms on the surface of dust grains.)  Solving the rate equations for dark hydrogen formation to determine whether it can cool the gas is complex.  Additionally, dark hydrogen is easily destroyed by radiation backgrounds with energies of $0.4 \alpha_x^2 m_c$ (and its catalyst H$^{-}$ is destroyed by $0.05 \alpha_x^2 m_c$ ones).  Indeed, once a very small number of baryonic stars formed in the Universe owing to molecular hydrogen cooling, the dissociating background from these stars is thought to destroy all molecular hydrogen in unenriched environments \citep{2000ApJ...534...11H, 2012ApJ...760....3M}.  Here we will consider the cases where molecules can cool the gas for all of our dark parameter space, noting that the case where they cannot was treated in the previous atomic cooling section.

\begin{figure*}
\includegraphics[width=\linewidth]{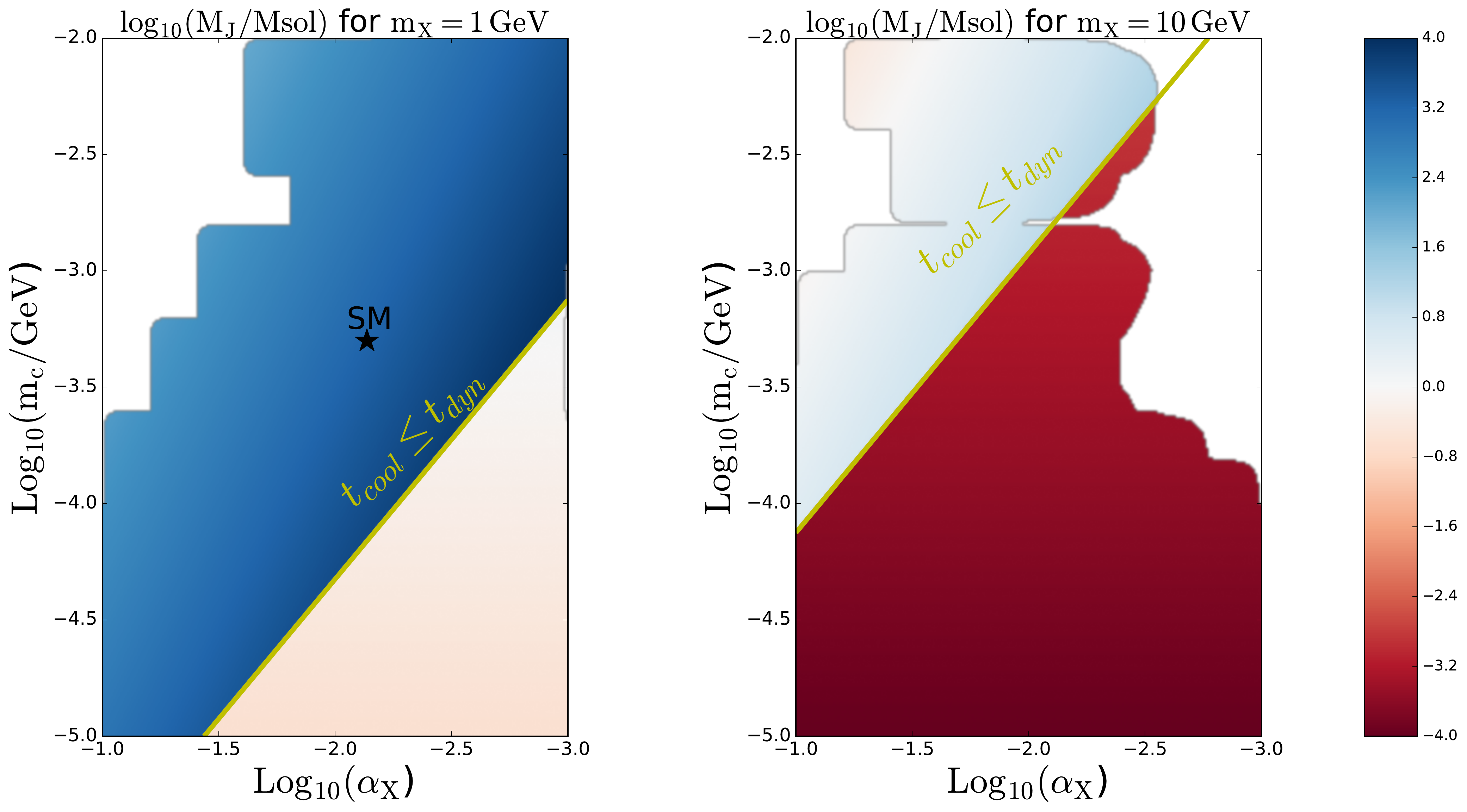}
\caption{Characteristic minimum clump mass in the efficient molecular cooling scenario, for models with $\epsilon=0.05$ and either $m_{X}=1~\GeV$ or $m_{X}=10~\GeV$.  The clump mass is set by the minimum Jeans mass a collapsing gas parcel attains, which is set by when cooling becomes inefficient.}
\label{fig:jeansmol}
\end{figure*}

In analogy to atomic gas, molecules can cool to $\sim \alpha_x^{2}m_{c}^2/m_{X}$ if it can cool via rotational transitions and till $\sim \alpha_X^{2}m_{c}^{3/2}/m_{X}^{1/2}$ through roto-vibrational transitions.  We assume the $\sim$ are equality for ensuing calculations. Rotational transitions are forbidden dipole transitions, whereas they are allowed at quadropole order. The rate is given by $\Gamma_{\rm quad} = \alpha^{7} m_{X}^{-5}m_{c}^6$. Roto-vibrational transitions can proceed through dipole radiation with the rate given by $\Gamma_{\rm dipole} = \alpha^{5} m_{X}^{-3/2}m_{c}^{5/2}$. Not all the parameter space can cool via rotational transitions because the spontaneous emission timescale can become longer than the dynamical time. In Fig \ref{fig:jeansmol} we show the parameter space that can cool via rotational transitions

For rotational transitions of H$_2$ molecules, the thermal equilibrium density can be calculated by equating the interaction rate to the quadropole radiation rate i.e. $\langle n_{\rm TE} \sigma v\rangle\,=\,\Gamma_{\rm quad}$ ,where $\sigma = \pi a^{2}_{0}$ and $a_{0}$ is the Bohr radius. 
The thermal equilibrium density for the rotational transitions is given by
\begin{equation}
\label{eq:thermal equilbriumrot}
n_{\rm TE}^{\rm rot} \sim 1\times 10^{4} {\rm cm}^{-3} \left(\frac{\alpha_{X}}{\alpha} \right)^{8} \left(\frac{m_c}{m_e} \right)^{7} \left(\frac{m_X}{m_p} \right)^{-4},
\end{equation}
where we have adjusted the coefficient from our approximate formula to match the precise value found in \citet{2002ApJ...564...23B} for the standard model, but used our scalings with $m_c$ and $m_X$. 
Using the final temperature reached by  purely rotational transitions $T^{\rm rot} = \alpha^{2}_{X}m_{c}\frac{m_{c}}{m_{X}}$ and the number density given in Eq. \ref{eq:thermal equilbriumrot} we get that Jeans mass scales as
\begin{align}
\label{eq:mjrot}
M^{\rm rot}_{J} \simeq 1500 M_{\odot} \times \left(\frac{\alpha_{X}}{\alpha} \right)^{-1}\left(\frac{m_c}{m_e} \right)^{-5/4}\left(\frac{m_X}{m_p} \right)^{-3/4}
\end{align}

Note that the above equation is just to indicate the scaling of the Jeans mass and is not applicable over the part of the parameter space where purely rotational cooling time is longer than the dynamical time of the halo and hence cannot cool the gas.

Once the spontaneous emission time becomes longer than the dynamical time, rotational cooling turns off and the gas reheats to temperatures where roto-vibrational transitions can cool the gas.  In this circumstance, the density at which roto-vibrational cooling becomes inefficient sets the Jeans mass at which fragmentation stops. The thermal equilibrium density for roto-vibrational transition is given by 
\begin{equation}
\label{eq:thermal equilbriumvib}
n_{\rm TE}^{\rm vib} \sim 2\times 10^{15} {\rm cm}^{-3} \left(\frac{\alpha_{X}}{\alpha} \right)^{6} \left(\frac{m_c}{m_e} \right)^{15/4} \left(\frac{m_X}{m_p} \right)^{-3/4},
\end{equation}

Using the final temperature of roto-vibrational cooling to be $T^{\rm vib} = \alpha^{2}_{X}m_{c}\sqrt{\frac{m_{c}}{m_{X}}}$ and density given by Eq. \ref{eq:thermal equilbriumvib} we get
\begin{align}
\label{eq:mjvib}
M^{\rm vib}_{J} \simeq 1 M_{\odot} \times \left(\frac{m_c}{m_e} \right)^{3/8}\left(\frac{m_X}{m_p} \right)^{-19/8}
\end{align}
Note that the above equation is to demonstrate scaling of the Jeans mass and is applicable only to the parts of the parameter space where purely rotational cooling cannot cool the gas. Since roto-vibrational transitions occur over much smaller timescales, this process allows the gas to cool to much larger densities which results into much lower Jeans mass.

In Fig. \ref{fig:jeansmol} we have plotted the expected Jeans mass over the parameter space that cools.  As long as the spontaneous emission timescale is less than the minimum dynamical time -- larger values of $\alpha_{X}$ and $m_{c}$ that we consider -- , cooling through purely rotational transitions sets the minimum Jeans mass (fragmentation mass).  Whereas for the lowest values of $\alpha_{X}$ and $m_{c}$, roto-vibrational transitions shapes the clump size.

Thus atomic and molecular cooling give us a wide range of Jeans masses ($10^{-3}M_{\odot}-10^{6}M_{\odot}$) over our parameter space. Molecular cooling generally gives smaller Jeans masses than atomic cooling.

\section{Constraints on dark baryons}
\label{sec:constraints}
In the previous sections, we calculated the mass of dark clumps assuming both atomic and molecular cooling. Since the Jeans mass sets the scale of fragmentation of the gas, we can assume that the mass of the dark clusters is approximately the minimum Jeans mass over its condensation history.  We note that it would be unsurprising for our characteristic mass to misestimate the actual mass by an order of magnitude.  We again take solace in that the large parameter space of models means that an order of magnitude shift in characteristic clump mass often shifts the ruled-out regions marginally.  In addition, one of our most stringent constraints turns out not to depend on the masses of the dark clumps.
\subsection{dynamical modeling (and strong lensing) mass estimates compared with mass-to-light ratio mass estimates}
\label{ss:gm}
This class of constraints is simpler as it does not depend on the additional step of estimating the mass of the dark clumps. Astronomers understand how to map stellar luminosity of a galaxy to its mass in stars at the factor of two level from studies that have modeled in detail populations of stars.  The factor of two estimated error stems from the amount the mass-to-light ratio vary between different models \citep{2013ARA&A..51..393C}.  Thus, we can observe the light of galaxies and use this to estimate the mass in stars.  These mass estimates can be compared with modeling that determines the enclosed mass, either by observing galactic rotation curves or by using strong gravitational lensing  (and on scales where standard dark matter should be a small contribution to the density).  Unlike the mass-to-light estimates, these other mass estimates are sensitive to the enclosed mass in atomic dark matter.
Thus, atomic dark matter in a bulge or disk cannot contribute more than the mass already in stars, as the mass in stars is approximately the error on mass-to-light inferences.  Let us first examine the case of the Milky Way.  Since only two percent of the Milky Way halo's mass is in the Milky Way galaxy itself \citep{2010ApJ...717..379B}, this constrains
\begin{align}
\label{eq:galmass}
\epsilon \times \left \{f_{\rm bulge}\left(\frac{r_{\rm MW}}{r_{\rm s}} \Big |_{\rm bulge}\right)^{3} + f_{\rm disk}\left(\frac{r_{\rm MW}}{r_{\rm s}}\Big |_{\rm disk}\right)^{2} \right \} \lesssim 0.02
\end{align}
where $r_{\rm MW}$ is the Milky Way stellar radius and $r_{\rm s}$ is the dark atom radius, either in the bulge or disk. We take $r_{\rm MW} = 3$~kpc. Our calculations show that the region of our parameter space where a disk forms has $r_{\rm s} \approx 2\,r_{\rm MW}$, although the difference from equality may owe to simplifications in our semi-analytic model (the assumed isothermal potential model overestimates the disk size and we did not model the distribution of halo spin parameters rather assuming the mean). However, in most of the remaining parameter space where a disk does not form and the cooling is efficient $r_{\rm s} < r_{\rm MW}$, leading to stronger constraints on $\epsilon$. The constraints on dark baryon parameter space assuming $\epsilon=0.05$ have been shown in Fig. \ref{fig:constraints}.

An equivalent constraint was published recently which constrained the surface mass density $\Sigma_{\rm MW}$ (and hence the galaxy mass) of our Milky Way disk by using stellar velocity data from Gaia sattelite \citep{Schutz:2017tfp}. However this study assumed that the entire parameter space of dark baryons that can cool forms a dark disk and contributes directly to $\Sigma_{\rm MW}$. Since this assumption is not valid for most of our parameter space we do not use this constraint.

In more massive halos than the Milky Way, the analogous constraint may be even stronger as the stellar-to-halo mass fraction is found to decrease with halo mass as $M_{\rm stars}/ M_{\rm DM} \approx 0.02 \left({M_{\rm DM}}/[10^{12} M_\odot] \right)^{-0.5}$ \citep{2010ApJ...717..379B}.  \citet{2015ApJ...800...94S} and \citet{2015MNRAS.446..493P} have constrained the masses in galaxies in the range of $10^{13}-10^{14.5} M_\odot$ from dynamical modeling combined with strong lensing mass estimates. We expect these measurements to be extremely constraining as, for example, the mass of a galaxy in a $10^{14}M_\odot$ halo is $2\times10^{-3}$ its halo mass, so naively a disk or bulge of $2\times10^{-3}$ in atomic dark matter would be ruled out.  However, deriving bounds from these more massive halos requires simulating a $10^{14}M_{\odot}$ merger tree to calculate bulge sizes, as in some parameter space a small fraction of the dark matter may reside in a bulge.
\subsection{constraints from ultra-faint dwarf galaxies}
\label{ss:uf}
Dynamical friction is a gravitational effect in which a heavier object deflects lighter ones, leaving a wake in its path, and the mass in this wake causes it to lose momentum and spiral towards the center of the potential. The momentum lost by the heavier object is then gained by lighter objects, `heating' them and causing their distribution to become more extended. In the parameter space in which dark clumps are considerably heavier than (baryonic) stars, dynamical friction acting on the dark clumps would increase the size of star clusters until they `dissolve' in their host galaxy.  Similarly, the stars in the galaxy themselves would experience the same `heating,' potentially increasing their half-light radius.  

Such constraints from dynamical friction heating are the strongest for the ultra-faint dwarf galaxies, satellites of Milky Way and Andromeda with halo masses of $\sim 10^{10}M_{\odot}$ -- the smallest galaxies and halos that are known.  Ultra-faint dwarf galaxies are the most dark matter-dominated systems in the Universe.  In particular, we can translate the MACHO and primordial black hole constraints from \cite{Brandt:2016aco} over to our dark baryons parameter space, which yields $$\epsilon\times f_{\rm cooled}\times M_{\rm ADM} \geq 10 M_\odot,$$ where $M_{\rm ADM}$ is the mass of atomic dark matter clumps and $f_{\rm cooled}$ is the fraction of gas cooled. The above bound comes from assuming the time to increase in half-light radii of the ultra-faint dwarf galaxies from 2 pc to 30 pc is larger than 10 Gyr (see Fig. 3 in \cite{Brandt:2016aco}).\footnote{Stronger bounds can be derived from the same paper using the evaporation of stellar clusters in the Eri~II galaxy.}  

\begin{figure*}
\includegraphics[width=\linewidth]{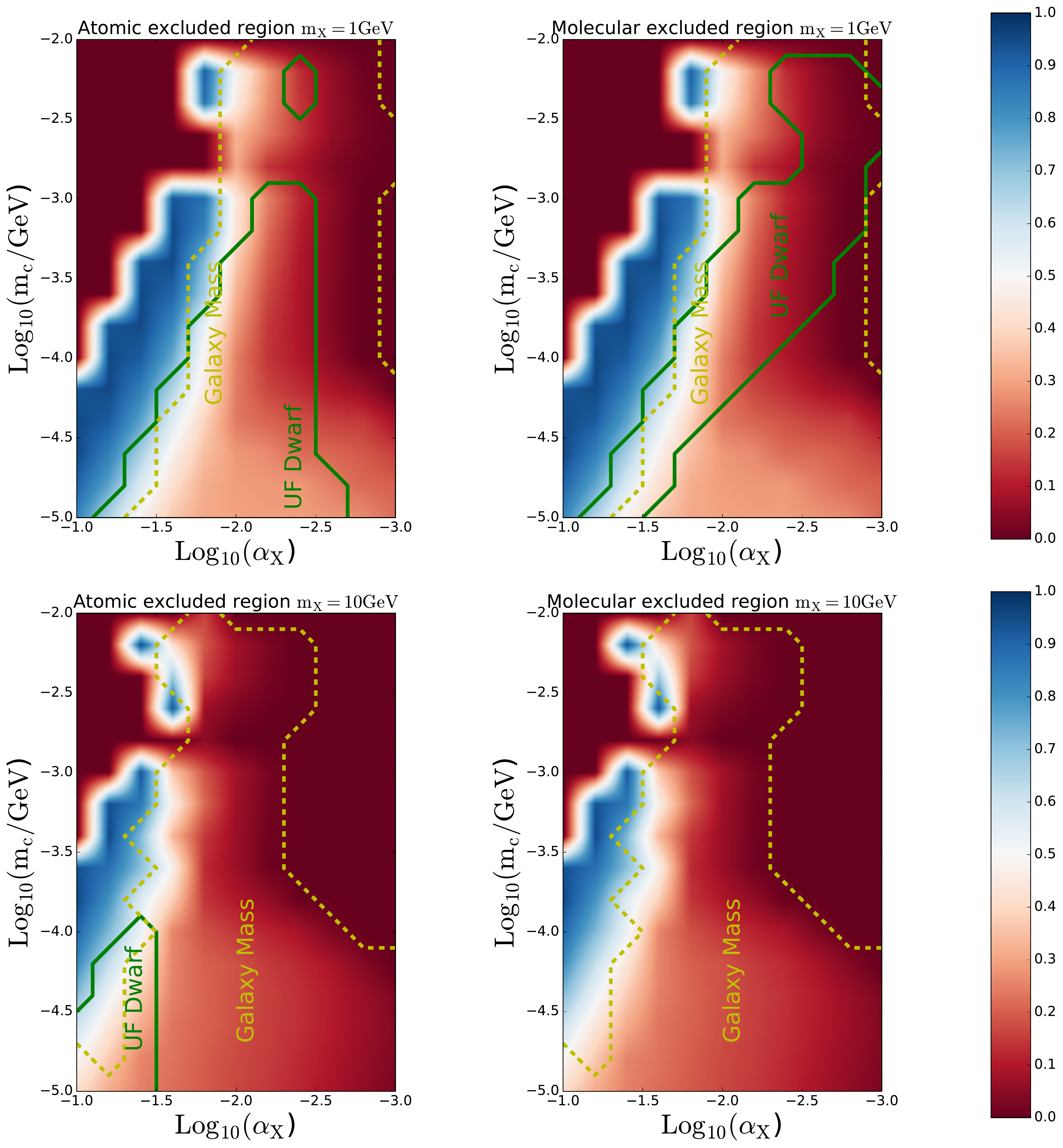}
\caption{Excluded regions for atomic dark matter models with $\epsilon=0.05$ and either $m_{X}=1~\GeV$ or $m_{X}=10~\GeV$ derived from different galaxy mass estimates (Section~\ref{ss:gm}) and from ultra-faint dwarf galaxies (Section~\ref{ss:uf}). The background shading shows the average bulge + disk mass fraction for $10^{12}M_{\odot}$ halos following the semi-analytic calculation described in Section~\ref{sec:galformation}.}
\label{fig:constraints}
\end{figure*}

This constraint from ultra-faint dwarfs is show in Fig. \ref{fig:constraints}. We did a merger tree simulation for $10^{10}M_{\odot}$ to estimate the parameter space where the `dark atoms' can cool and form dark clumps. We have not included the fact that in parts of the parameter space with large spheroid fractions the dark clumps will be much more concentrated near the center of dwarf galaxy increasing (by a huge factor) the rate of `heating' caused by their dynamic friction.  Nevertheless, even with this extremely conservative assumption, much of the parameter space in which $ M_{\rm ADM} >200 M_\odot$ is ruled out.  

\subsection{microlensing constraints}
The dark clumps will act as gravitational lenses to stars, leading to enhancements in flux of these stars that are of the order one when the star passes in projection within the Einstein radius of a lens, $R_E$.  The timescale for this enhancement is $t_{\rm ml} \sim R_E/v$, which corresponds to of the order $100~$days for a solar mass star, assuming that the the relative velocity of the star with respect to the lens is a Milky Way-like $v=200~$km~s$^{-1}$.

On a halo scale, the MACHO survey offers the best constraints on dark clumps \citep{2000ApJ...542..281A}.  This survey searched for stellar lensing events towards the Large Magellanic Clouds.  MACHO constrains dark clumps distributed like an NFW halo to be $<10\%$ of the dark matter for masses in the range $10^{-7}-10^{-3} M_\odot$ \citep{1998ApJ...499L...9A}, and $<40\%$ of the dark matter for $M_\odot = 1-30$ \citep{2015MNRAS.446..493P,2015ApJ...800...94S}, and somewhat weaker constraints between these ranges (although there are also more lenses than expected in this range, allowing for tens of percent of the dark matter).   Unfortunately, because we consider $\epsilon < 0.05$ and in most of our cooling scenarios the dark clumps are more concentrated than the halo dark matter, which suppresses the lensing efficiency by the fractional distance to the lens relative to a lens at half the source distance (for us a factor of $>10$), halo-scale microlensing does not constrain viable parameter space for atomic dark matter.

Galactic microlensing surveys, which instead use source stars within the Milky Way, provide more of a constraint on our model. The OGLE survey detects thousands of lensing events, with the histogram of event durations peaking at $30~$days, as expected for stars -- few tenths of solar mass objects  \cite{2015ApJS..216...12W}.  Less than a tenth of events that occur have $3$ day timescale,  which would correspond to $M_{\rm ADM} \sim 10^{-3}M_\odot$.   Thus, the number of solar mass objects relative to stars should be $\rho_{\rm ADM} = 0.1 \left([0.1 \Msun]/{M_{\rm ADM}}\right)^{1/2} \rho_{*}$, where the  factor in parentheses accounts for the higher rate of lower mass lenses and $\rho_*$ is the number density in stars and we have taken their characteristic mass to be $0.1 \Msun$.  Thus, for $M_{\rm ADM} \sim 10^{-3}M_\odot$, $\rho_{\rm ADM} < \rho_{*}$.   Finally, note that the mass in stars in Milky Way-like galaxy is estimated to be $\approx 0.02 M_{\rm DM}$ from galaxy--halo abundance matching \citep{2010ApJ...717..379B}.  Thus, if all the dark clumps are in a disk, using that $\rho_{*} \propto 0.02 M_{\rm DM}/[0.1 \Msun]$ and using $\epsilon=0.05$ we get
\begin{align}
f_{\rm disk} \lesssim 4\times 10^{-3} {\rm ~~~~~for~~~~~} M_{\rm ADM} \sim 10^{-3}M_\odot.
\end{align}
The constraints are less stringent on the case where the lensing is from a bulge, where since a bulge is more extended, within the plane of the galaxy $l_{\rm disk}/R_{\rm bulge} \sim 0.1$ and hence
\begin{equation}
f_{\rm bulge} \lesssim 4\times 10^{-2} {\rm ~~~~~for~~~~~} M_{\rm ADM} \sim 10^{-3}M_\odot.
\end{equation}
This neglects that some source stars that have been surveyed are bulge stars, which will be more sensitive to bulge atomic dark matter.

These constraints are extremely powerful and are likely to rule out much of the dark atom parameter space for $m_{X}=10~\GeV$ in the the molecular cooling scenario, where we find $M_{\rm ADM}\sim 10^{-3}\Msun$ (and roto-vibrational cooling is dominant). However, it does not rule out any of our dark atom parameter space for $m_{X} = 1~\GeV$, since the characteristic mass scale there is $> 10^{-1}\Msun$.

The above estimates are mostly for illustration, and we do not include the microlensing constraints in \ref{fig:constraints}. In order to derive proper constraints one would need to take into account the correct spheroidal radius of our `dark galaxy,' which can change the distances between lens and source and the correct modeling of the spread in the lensing timescales.

Future shorter cadence microlensing surveys should be powerful for constraining the sub-solar mass atomic dark matter parameter space, such as the upcoming NASA mission WFIRST.\footnote{\url{https://wfirst.gsfc.nasa.gov/}}  WFIRST will be sensitive to lenses with masses as small as $10^{-8} M_\odot$ \citep{2012MNRAS.423.1856S}, providing extremely sensitive constraints on $\epsilon$ in our models, especially in the regions where roto-vibrational molecular cooling sets the mass of stars.  The limitation for such constraints comes from not knowing how much mass resides in planets and asteroids in the Galaxy.  Models predict planet/asteroid mass densities that are at least two orders of magnitude down from the mass in stars (and their masses are concentrated at certain scales).  Thus, future microlensing surveys will dramatically constrain $\epsilon$ for our parameter space with $M_{\rm ADM} \lesssim 10^{-3} M_\odot$.

\section{Conclusions}
This paper explored the constraints on a fraction of the dark matter being two darkly charged particles with a large mass ratio and a massless force carrier.  In such models, the dark matter can cool and exhibit interesting dynamics.   Previous studies assumed that, in parameter space where the dark matter can cool, it would form a disk \cite[e.g.][]{2013PDU.....2..139F}.  This paper investigated this assumption by running more detailed `semi-analytic galaxy formation models', a technique borrowed from studies of galaxy evolution.  These models track the distribution of atomic dark matter by considering the timescales for energy exchange, cooling, and dynamical friction on top of realizations for the growth and merger histories of dark matter halos.  By running a large grid of such calculations, we found that -- if the atomic dark matter can cool -- only in a narrow range of parameter space does it cool late enough for a disk to remain.  Rather, in most of the parameter space, the cooling occurs at relatively high redshifts.  As cooling is a runaway process, the end result will be dense clumps that behave collisionlessly when galaxies merge.  Thus, once the atomic dark matter has cooled and merged with a dark disk of comparable mass, the resulting gravitational torques will destroy the dark disk, resulting in a spheroidal bulge on galactic scales.  Additionally, when smaller systems merge with larger ones, there is often insufficient time to spiral to the center by dynamical friction and, indeed, we find for a Milky Way-like halo that most of the atomic dark matter remains distributed on the $200~$kpc scale of the dark matter halo.

The conclusion that $z=0$ atomic dark matter disks are unlikely is not surprising when considering how the standard model baryons behave.  Even though the standard model baryons can cool in $10^8-10^{12}M_\odot$ halos on a cosmologically short time, stellar feedback (driven largely by supernovae) dramatically suppresses the formation and growth of galaxies, mimicking the effect of much longer cooling times.  Indeed, the Milky Way galaxy contains only $\sim 10\%$ of the baryons that should be present given its halo mass because of feedback, and this fraction dramatically decreases in smaller mass halos as feedback becomes more effective.
  The low masses of galaxies in smaller halos results in, even after a merger that disrupts their disks, significant growth of a new disk as uncooled baryons cool in the merged system.  
   However, without stellar feedback, a large fraction of the baryons would have turned into stars in disks at much earlier times, and later merging and disruption would result mainly in spheroidal bulges or more diffuse distributions. Indeed, stars distributed in bulges or on the halo-scale are exactly the picture for the distribution of stars in more massive halos than the Milky Way's.  

Thus, previous constraints, which assumed that all the atomic dark matter resides in a disk, do not apply to most atomic dark matter parameter space.  Still, we found that there were interesting constraints on the dark atom scenario.  Indeed, much of the parameter space for atomic dark matter is ruled out if five percent of the dark matter is atomic ($\epsilon = 0.05$) by mass-to-light ratio measurements coupled with dynamical modeling of the Milky Way rotation curve.  A dark bulge can also be constrained by MACHO constraints as cooling dark atoms would condense into dense clumps analogous to how the baryons fragment on solar mass scales.  The size of fragments is set by the scale at which the gas can fragment -- the distance a sound wave travels in a collapse time -- reaches a minimum, which happens when cooling becomes inefficient. We argued that this scale is likely set by the onset of free-free opacity, unless the dark matter is able to form sufficient molecules to cool.  Then, this scale is set by when the rotational or vibrational transitions reach thermal equilibrium.  
 These scales enable rough estimates for the size of dark matter fragments.  We found that much of the parameter space for substantial dark fragments is constrained by Galactic microlensing and the small half-light radii of ultra-faint dwarf satellite galaxies. While current microlensing constraints only constrain a fraction of this allowed parameter space with $\sim 10^{-3} M_\odot$, the WFIRST satellite should be able to rule out dark clumps masses of $10^{-8}-10^{-4} M_\odot$ even for the case of $\epsilon \sim 10^{-3}$.  More massive dark clumps are ruled out by ultra-faint dwarf satellites for $\epsilon \sim 0.05$.

A concern is whether our assumption of no feedback in the dark sector is reasonable.  We note that small tweaks to the standard model would turn off stellar feedback processes -- a smaller fine structure constant would make all stars red \citep{carter74}, eliminating radiative feedback, and Type II supernovae likely would not explode with even minor tweaks to any standard model force.  A minimal atomic dark matter model does not have `dark nuclear physics', such that the standard mechanisms that power feedback are not present. A natural extension of the atomic dark matter that would feedback would be to allow dark matter to annihilate at the center of our clumps, powering dark stars.  If there is a coupling to the SM, these stars could manifest as Galactic point sources \cite{Agrawal:2017pnb}.

Our calculations have relevance for considerations of how tuned our universe is for life.  Previous anthropic studies that focused on cosmology and astrophysics have analyzed the apparent fine-tuning of the cosmic initial conditions to have nonlinear structure on galactic scales \citep{RevModPhys.61.1, 1998ApJ...499..526T, 2007MNRAS.379.1067P} and of the fundamental constants to have stars and supernovae \citep{1979Natur.278..605C, 2006PhRvD..73b3505T, 2017APh....87...40A, 2016JCAP...02..042A}.  Our study shows that the space in between structure formation and star formation also requires tuning to produce (1) cooled gas,  (2) well-ordered galactic disks of stars, and (3) Chandrasekhar mass stars as is necessary for supernovae. We found that in order to gas to cool, the dark electron must be $< 10^{-2}$ MeV mass, and the proton must be not much heavier than $\sim 10 ~\GeV$ (to prevent large virial temperature and large equlibration times with electrons) or much lighter than $\sim 1 ~\GeV$ (to prevent the virial temperature from being too small for the gas to ionize and cool). Finally, our results highlight the importance of stellar feedback in shaping our galactic disk. Studies have found that living at the edge of an ordered spiral galaxy may enhance habitability \citep[e.g.][]{2011AsBio..11..855G}.

While there are many deficiencies with this work (such as determining whether molecules would form, the exact spectrum of dark fragments, and whether feedback operates) remedying these deficiencies could be challenging (as all of these example are likely coupled).  However, if an excess of microlensing events is discovered, a plethora of unexplained $\gamma$-ray point sources, or too many massive merging black holes are found by gravitational wave observatories (perhaps with masses that were not anticipated for stellar remnants;\citep[]{2016PhRvL.116t1301B}), such observations could motivate additional effort towards understanding atomic dark matter.\\

MM thanks the Alfred P. Sloan foundation and the Institute for Advanced Study visiting faculty program. We would also like to thank Anson D'Aloisio,  Vid Irsic and Neal Dalal for helpful conversations.
\appendix
\section{Atomic Cooling}\label{atomiccool}
Atomic cooling happens when a (dark) electron loses energy after colliding with an atom via collisional ionization or collisional excitation.  The cross-section for collisional ionization is given by \citep{PhysRevA.50.3954,Rosenberg:2017qia}
\begin{align}
\sigma_{\rm ion}\, &=\, \frac{8\pi a^{2}_{0}}{\frac{K}{B}+\frac{3}{2}}\Bigg[ \frac{\log\left(\frac{K}{B}\right)}{2}\left(1-\left(\frac{K}{B}\right)^{-2}\right)
\\
&+\left(1-\left(\frac{K}{B}\right)^{-1}-\frac{\log\left(\frac{K}{B}\right)}{1+\frac{K}{B}}\right)\Bigg],
\end{align}
and that of collisional excitation of the Ly$\alpha$ line is given by \citep{RevModPhys.43.297}
\begin{equation}
\sigma_{\rm coll}\,=\, \frac{4\pi a^{2}_{0}\,B}{K+\frac{7\,B}{4}}\left[0.55\left(-0.89+\log\left(\frac{4\,K}{B}\right)+0.208\frac{B}{K}\right)\right]
\end{equation} 
where $a_{0}$ is the Bohr Radius, $B$ is the binding energy and $K$ is the kinetic energy of the incoming electron. Integrating these cross section over thermal distribution of electrons and multiplying by the number of electrons yields
\begin{equation}
\label{gatomic}
\Gamma_{\rm atomic}\,=\,x\left(1-x\right)n_{c}(z)\left(\langle\sigma v\rangle_{\rm ion}+\frac{3}{4}\langle\sigma v\rangle_{\rm coll}\right),
\end{equation}
where we have ignored the cross section for the collisional excitation into other states and $\langle\sigma v\rangle_{\rm ion},~\langle\sigma v\rangle_{\rm coll}$ are the velocity-averaged collisional ionization and collisional excitation cross sections, respectively.
 The only undetermined part here is the ionization fraction $x$, which is a function of temperature and can be calculated by comparing recombination rates and ionization rates:
 \begin{equation}
x = \frac{\langle\sigma_{\rm ion} v\rangle}{\langle\sigma_{\rm ion} v\rangle+2\,\langle\sigma_{\rm rec} v\rangle}.
\end{equation} The velocity averaged recombination cross section can be calculated by using Milne relations and is given by (see Ch.10 in \cite{ryb})
\begin{equation}
\langle\sigma_{\rm rec} v\rangle = \pi a^{2}_{0}\frac{64}{3\sqrt{3}}\left(\frac{m_{c}}{2\pi T}\right)^{3/2}\alpha^{7}_{X}\left(e^{\frac{B}{T}}\int^{\infty}_{\frac{B}{T}}\frac{e^{-t}}{t}\right).
\end{equation}

\section{Merger Tree Algorithm}
\label{mergeralgo}
In this Appendix we will describe the method and pseudocode for the creation of a merger tree and the techniques and pseudocode we use to go through the merger process to study galaxy formation. We will start by describing the merger tree formation process.  Our algorithm is largely the same as that of \cite{lacey,Cole:2000ex,lacey}; please consult this reference for a more thorough description.

\subsection{Merger Tree Creation}
\label{ap:mergertreecreation}
 The matter overdensity field in our universe is to excellent approximation at Gaussian with standard deviation as a function of halo mass $M$ today as
\begin{align}
S(M) \equiv \sigma^{2}(M) = \int \frac{k^{2}dk}{2\pi^{2}}P(k)\left(\frac{j_{1}(kR)}{kR}\right)^{2},
\end{align}
  where $P(k)$ is the $z=0$ linear matter power spectrum, $j_{1}$ is the spherical Bessel function, and $R= (3 M/4\pi)^{1/3}$ is the Lagrangian radius.  

To determine the merger history of a single dark matter halo at $z=0$ there is a well-tested prescription called a `halo merger tree' that can be derived from the model described in \citep{Cole:2000ex} by solving simple diffusion equation, with ``time'' being the smoothing scale $R$ and ``position'' being the variance of the density, with an absorbing boundary.  The diffusion describes how the overdensity of different regions varies as a function of $R$: When an overdensity hits the absorbing boundary (also called the ``barrier''), which evolves with redshift according to Eq.~\ref{eqn:deltac}, it has collapsed into a halo in this model. The probability that a halo of mass $M_{1}$(or a corresponding variance of the overdensity $S_{1}$) at time $\omega(z_{1})=\omega_{1}$ will contribute to a halo of size $M_{2}$ ($S_{2}$) at  
time $\omega(z_{2})=\omega_{2}$ is given by 
\begin{align}
\label{eq:massfunc}
f(S1,\omega_{1}|S2,\omega_{2}) = \frac{\omega_{1}-\omega_{2}}{\sqrt{2\pi (S_{1}-S_{2})^{3}} } e^{-\frac{(\omega_{1}-\omega_{2})^{2}}{2(S_{1}-S_{2})}} .
\end{align} 
This expression gives the distribution a halos at $z_1$ that merge into a halo of size $S_2$  at redshift $z_2$ and, hence, can be used to generate a merger tree.

Equation \ref{eq:massfunc} describes the distribution of the fraction of mass in halos of mass $M_{2}$ at time $t_{2}$ that at an earlier time, $t_{1}$, was part of a halo of mass $M_{1}$. The expression can be used to generate a merger history of our halo by drawing from this probability distribution: Imagine we take an infinitesimal time step $dt_{1}$; the number of subhalos of mass $M_{2}$ that $M_{1}$ fragments into is given by 
\begin{align}
\label{eq:numassfunc}
\frac{dN}{dM_{1}}\,=\,\frac{df_{12}}{dt_{1}}\frac{M_{2}}{M_{1}}dt_{1},
\end{align}
where
\begin{align}
\label{eq:numassfunc}
\frac{df_{12}}{dt_{1}} \Big |_{t_{1}=t_{2}}\,=\,\frac{1}{(2\pi)^{1/2}(S_{1}-S{2})^{3/2}}\frac{d\delta_{c1}}{dt_{1}}\frac{dS_{1}(M)}{dM}.
\end{align}
It follows that the probability that the halo $M_{2}$ splits into a halo of mass $M_{1}$ such that $M_{\rm res}<M_{1}<M_{2}/2$ is given by
\begin{align}
\label{eq:probsplit}
P(M_2) = \int^{M_{2}/2}_{M_{\rm res}}\frac{dN}{dM_{1}}dM_{1}
\end{align}
where $M_{\rm res}$ is the resolution mass below which our simulation does not resolve individual halos.  In order to take into account halos below $M_{\rm res}$, the fraction of mass in these halos that is `accreted' is given by
\begin{align}
\label{eq:accretefrac}
F(M_1, M_2) = \int^{M_{\rm res}}_{0}\frac{dN}{dM_{1}}\frac{M_{1}}{M_{2}}dM_{1}.
\end{align}

The particular specifications we use for our merger tree calculations are as follows. We generate the merger tree between $z_{i} = 0$ to $z_{f} = 10$ and with mass resolution $M_{\rm res} = 3\times 10^{7} M_{\odot}$. We then divide the time $t_{i}$ to $t_{f}$ into logarithmic time steps. The number of time steps is chosen such that $P \ll 1$, making it unlikely there is more than one splitting of a halo at a given timestep. For the purposes of generating our merger tree, we used $t_{\rm steps} = 7000$.   For most of our calculations, our starting ($z=0$) mass is $M_{2} = 10^{12}M_{\odot}$. Pseudocode for the merger tree algorithm is given in Fig. \ref{fig:mergeralgo}. We use the merger tree generated by the above algorithm to study galactic-scale structure of atomic dark matter.

\begin{figure*}
  \caption{Merger tree algorithm.  See appendix~\ref{ap:mergertreecreation} for relevant definitions.}\label{fig:mergeralgo}
  \begin{algorithmic}[1]
    \Procedure{Merger}{$10^{12}M_{\odot}$}\Comment{Merger tree function evaluated for halo mass of $10^{12}M_{\odot}$ }
    \For{$i = 0$ to ${nsteps}$}
            \State $M_{1} = [] $\Comment{no daughter halos at the beginning}
            \State $index = []$\Comment{keeps track of the halo index that splits}
                        \For{$j$ in $M_2$}\Comment{scan the parent halos}
							\If {$rand < P(M_{2}\left[j\right])$}\Comment{halo splits if uniform random number meets condition}
                			\State $M_{d1} = rand(M_{\rm res},M_{2}\left[j\right]/2)$\Comment{randomly pick daughter halo mass between $M_{\rm res}$ and $\frac{M_{2}}{2}$ consistent with \ref{eq:massfunc}}
                			\State $M_{d2} = M_{2}\left[j\right](1-F(M_{2}\left[j\right]))-M_{d1}$\Comment{reduce the accreted fraction from parent halo and subtract the split halo mass}
                			\State $M_{1} = M1.append(M_{d1},M_{d2})$\Comment{append the daughter halos to array keeping track of daughter halos}
                			\State $index = index.append(j)$\Comment{store index the number of the parent halo that split}
                			\ElsIf {$M_{2}\left[j\right] > M_{\rm res}$}\Comment{If $M_{2}$ is above the resolution mass and does not split}
                			\State $M_{1} = M1.append(M_{2}\left[j\right](1-F(M_{2}\left[j\right])))$\Comment{subtract the accreted mass}
		 			        \Else\Comment{if $M_{2}\left[j\right] < M_{\rm res}$}
                			\State $do\,nothing$ \Comment{If $M_{2}\left[j\right]\leq M_{\rm res}$ the halo stops evolving}
            				\EndIf
            			\EndFor
            \State $M_{2}=M_{1}$\Comment{Set the daughter halo array to be the parent halo array for next step}	
    \EndFor
    \EndProcedure
  \end{algorithmic}
\end{figure*}

\subsection{Galaxy Formation}
Now that we have constructed the merger tree, we use it to follow the cooling and contraction of dark atoms and, hence, their galactic-scale structure. We start at the top of our merger tree at $z = 10$. As we walk the tree towards $z = 0$, we follow the halos as their atomic dark matter cools and as the halos merge with each other. The rules we apply follow the semi-analytic galaxy formation algorithm of \cite{Cole:2000ex}, except that we use the timescales relevant to the atomic dark matter model of interest.  In what follows, we describe these rules.

First, we trace each halo from its smallest progenitors through all mergers it experiences. The lifetime of a halo is defined as the time it takes for the halo to double in mass.  Mass doubling can be driven by accretion, by merging with a larger halo, or by a combination of accretion and mergers.  During a halo's lifetime, we take its properties ($T_{\rm vir}$, $t_{\rm dyn}$, etc.) to be those at formation.  Halos whose mass increases above the resolution mass, $M_{\rm res}$, as we follow the merger tree are defined to have formed at the earliest time at which its mass satisfies $M>M_{\rm res}$. 

Next, at every time step in which two halos merge, we check if all the halos were able to cool during their lifetime.  The cooled dark atoms have three components: a halo component (which has not contracted), and a disk or bulge component (that is comparable in size to a galaxy). When the gas in a halo cools, it first ends up contributing to the disk.  We keep track of the mass in all three components for each halo.  

At every time step in which two halos merge we check if all the halos at that time step were able to cool during their lifetime i.e. since the time the new halo they are associated with was defined to have formed. Every galaxy has two components i.e. the disk and the bulge. When the gas in a halo cools it ends up contributing to the disk of the galaxy . At all the time steps we consider, we keep track of fraction of mass in the disk and bulge of the central galaxy and the total fraction of baryons that have cooled.

When a new halo forms we list all its satellite halos i.e. the halos that merged with it during its formation. We only keep track of the satellite halos that merged with it i.e. we don't care about the substructure of the merging satellite halos. The merging process is governed by dynamical friction. Dynamical friction can be thought of in the following terms. The satellite halo moves through a cloud of smaller mass object gravitationally dragging them alongwith producing a `gravitational wake'. The gravitational force of the objects in the `gravitational wake' on the satellite halo causes it to slow down and merge into the central halo. The merger time is given by \cite{Cole:2000ex}
\begin{align}
\label{eq:tmerge}
t_{\rm mrg} = \Theta \, t_{\rm dyn}\frac{0.3722}{\ln\left(M/M_{\rm sat}\right)} \frac{M}{M_{\rm sat}}
\end{align}
where $M$ is the total mass of the new halo, $M_{\rm sat}$ is the mass of the merging satellite galaxy including the mass of the halo it is in, $t_{\rm dyn}$ is the dynamical time given in Eq. \ref{eq:tdyn}, and $\Theta = 10^{-0.14}$ is a orbital parameter which parametrizes the energy and orbital angular momentum of the merging satellite (See \cite{Cole:2000ex}) .\footnote{$\Theta$ has a log-normal distribution with mean of -0.14 and standard dev of 0.26 in base $10$. Here we have used only the mean.}
The atomic dark matter bulges/disks that do not merge with the central bulge/disk in the lifetime of the halo become the satellite galaxies of the new halo. Every time a satellite merges with a central disk, the disk is disrupted if the masses of the two ``galaxies'' are comparable so there mutual gravitational forces destroy the disk.  We take $0.5 M_{\rm cen} \leq M_{\rm sat} \leq 2 M_{\rm cen}$ for disruption to occur as motivated by numerical studies, but we find our results are not significantly changed if instead disruption occurs when $0.3 M_{\rm cen} \leq M_{\rm sat} \leq 3 M_{\rm cen}$.  The end result of the disrupted merged system is a spheroidal bulge.  Otherwise, the disk is not destroyed and the satellite contributes to the ``galaxy'' bulge.  We also track the radii of the disk when they form and through their mergers. The radius of a disk or bulge  within a halo is given by
\begin{align}
\label{eq:rgal}
r_{\rm gal} = \frac{\lambda}{\sqrt{2}}r_{\rm vir}\left(M,z\right),
\end{align}
where $\lambda=0.039$ is the mean spin parameter found in cosmological simulations and $r_{\rm vir}\left(M,z\right)$ is the virial radius of the halo with mass $M$ and the redshift $z$ at which the gas cools to form a ``galaxy'' \citep{BarkanaLoeb2001}.\footnote{While halos show an approximately log-normal distribution of spin parameters, $\lambda$, with half a decade standard deviation, we model all halos to have the mean $\lambda$.} In case any additional gas accretes onto an existing disk at a lower redshift, we take the mass-weighted average of the two disk radii to find the new $r_{\rm gal}$. In the event of a merger, the new radius can be calculated as
\begin{align}
\label{eq:rnew}
r_{\rm gal,new} = \frac{\left(M_{1}+M_{2}\right)^{2}}{\frac{M^{2}_{1}}{r_{1}}\,+\,\frac{M^{2}_{2}}{r_{2}}+\frac{2M_{1}M_{2}}{r_{1}+r_{2}}}.
\end{align}

Following the above strategy yields the distribution of atomic dark matter (in the halo, bulge and disk) in a present day $10^{12}M_{\odot}$ halo as well as in its daughter halos.

\citestyle{plain}
\bibliography{References}

\end{document}